\documentclass[useAMS,usenatbib,usegraphicx]{mn2e}
\usepackage{color} %  Para usar colores en el texto

\def\kmskpc{{\rm\,km\,s^{-1}{kpc}^{-1}}}

\def\mathnew{\mathsurround=0pt}
\def\simov#1#2{\lower .5pt\vbox{\baselineskip0pt
    \lineskip-.5pt\ialign{$\mathnew#1\hfil##\hfil$\crcr#2\crcr\sim\crcr}}}

\def\'#1{\ifx#1i{\accent"13\i}\else{\accent"13#1}\fi}

\def\aap{{A\&A}}
\def\aaps{{A\&AS}}

\def\aj{{AJ}}
\def\apj{{ApJ}}
\def\apjl{{ApJL}}

\def\mnras{{MNRAS}}

\def\rmxaa{{Rev. Mex. Astron. Astrofis.}}
\def\pasp{{PASP}}
\def\msais{{MSAIS}}

\title[Resonant Trapping in the Galactic Disc and Halo and its 
Relation with Moving Groups]{Resonant Trapping in the Galactic Disc
and Halo and its Relation with Moving Groups}

\author[Autores] {Moreno, E.$^{1}$\thanks{E-mail:
edmundo@astro.unam.mx}, Pichardo, B.$^{1}$, Schuster, W. J.$^{2}$,
\\ $^{1}$Instituto de Astronom\'ia, Universidad Nacional
Aut\'onoma de M\'exico, Apdo. Postal 70-264, Ciudad Universitaria
D.F. 04510, M\'exico\\ $^2$Instituto de Astronom\'ia, Universidad
Nacional Aut\'onoma de
M\'exico, Apdo. Postal 106, 22800 Ensenada, B.C.,
M\'exico}

\begin{document}

\maketitle

\label{firstpage}

\begin{abstract}

With the use of a detailed Milky Way nonaxisymmetric potential,
observationally and dynamically constrained, the effects of the bar
and the spiral arms in the Galaxy are studied in the disc and in the
stellar halo. Especially the trapping of stars in the disc and Galactic
halo by resonances on the Galactic plane, induced by the Galactic bar,
has been analysed in detail. To this purpose, a new method is presented
to delineate the trapping regions using empirical diagrams of some
orbital properties obtained in the Galactic potential. In these
diagrams we plot in the $inertial$ Galactic frame a characteristic
orbital energy versus a characteristic orbital angular momentum, or
versus the orbital Jacobi constant in the reference frame of the bar,
when this is the only nonaxisymmetric component in the Galactic
potential. With these diagrams some trapping regions are obtained in
the disc and halo using a sample of disc stars and halo stars in the
solar neighbourhood. We compute several families of periodic orbits on
the Galactic plane, some associated with this resonant trapping. 
In particular, we find that the trapping effect of these resonances
on the Galactic plane can extend several kpc from this plane, trapping
stars in the Galactic halo. The purpose of our analysis is to
investigate if the trapping regions contain some known moving groups
in our Galaxy. We have applied our method to the Kapteyn group, a
moving group in the halo, and we have found that this group appears
not to be associated with a particular resonance on the Galactic
plane.

\end{abstract}
 

%%%%%%%%%%%%%%%%%%%%%%%%%%%%%%%%%%%%%%%%%%%%%%%%%%%%%%%%%%%%%%%%%%%%%%%
\section{Introduction}
\label{sec:intro}

Since the discovery of stellar associations or streams, now also called
`stellar moving groups' \citep[and references therein]{P69,ES59,E77,
E90,E96a,E96b,WR38,R49,SM93,M94,MHM96}, diverse theories to explain
them have emerged, from simple disruptions of open clusters by Galactic
tidal effects, to cosmological origins, such as dwarf satellite
galaxies or globular clusters \citep{PWG09}, to perturbations by
resonances with the Galactic bar and/or spiral arms \citep{DE00,F01,
AVP09,AFR11,QDV11}, and also triaxially shaped dark matter haloes
\citep{RVP12}. Most likely, all these processes together contribute to
form the multiple moving groups seen today in the Galaxy, but is
there one more effective than the others, should we be seeing more
groups produced by intrinsic secular processes than the ones produced
by disrupted satellites or clusters, or can we identify clearly what
phenomenon produces which particular group? Hopefully soon, these and
many other questions will be better answered with the impending arrival
of the great new large surveys (such as Gaia, SLOAN/APOGEE, the
Geneva-Copenhagen Survey, RAVE, SEGUE, etc.) joined to the best
theoretical tools to elucidate the different processes.

Meanwhile, one of the intricate problems to solve today is the
detection of these groups. Some of the best known methods to identify
moving groups in the literature are those using kinematic diagrams,
such as the Bottlinger diagram \citep{B32,TW53} in the U,V,W space,
and Color--Magnitude (CM)
diagrams \citep{E58,E59,ES59,E60a,E60b,E65b,E70,E71a,E71b,E74,E76,
E77,E83,E90,E96a,E96b,SSC12}. Also the method of \citet{LBLB95} and of
\citet{LB99} that analyses positions and velocities of globular clusters
and satellite galaxies, searching for halo fossil tidal streams that
mark orbits followed by satellites merging with the Galaxy. Another 
method is the one known as Great Circle Cell Counts (GC3) by
\citet{JHB96}, that makes use of the fact that orbits dominated by the
spherical outer halo conserve approximately their orbital plane
orientation and leave behind `great circles' on the sky. And, methods
to search for moving groups in the disc, that work in velocity space,
considering that all the stars in a given moving group move toward the
same direction \citep[e.g.][]{HA99,dB99}. A method based on the 
lumpiness in integrals of motion has been employed by \citet{HZ00}
to study the merger history of our Galaxy. For a review of methods,
see this last reference. 

Previous work has demonstrated the influence of galactic bars on
resonant trapping of orbits on the Galactic plane and out of it.
For example particles on bar resonances can exchange or transfer
angular momentum to particles in the dark matter halo, driving 
dynamical evolution (\citep{LBK72,TW84,A02,MV06,S14,CK07} and
references therein). Although N-body simulations will probably be the
best test-bed models to produce studies like the one presented here,
it is still intricate to extract information from resonant particles
because of their lack of resolution (\citep{WK07a,WK07b}). Therefore,
we employ in this work a very detailed Milky-Way steady model that is
described in Section \ref{sec:model}. 

Assuming that a large fraction of moving groups can be explained by
intrinsic properties of the Milky Way Galaxy, such as its
nonaxisymmetric potential structure, in this work we analyse the
trapping of stars by resonances on the Galactic plane produced by these
nonaxisymmetric components, and our aim in this and in a second paper
is to investigate the possible relation of these trapping regions with
some known moving groups in our Galaxy. For this purpose, we have 
developed a new straightforward tool to analyse the resonant trapping.
The method requires the computation of stellar orbits in a
nonaxisymmetric Galactic potential, and makes use of diagrams of a
`characteristic' orbital energy versus a `characteristic' orbital
angular momentum, or versus the orbital Jacobi constant if the Galactic
potential employs only the bar in its nonaxisymmetric components.

A sample of halo stars and disc stars in the solar
neighbourhood is employed to show how the proposed method works.
Our Galactic potential uses mainly the bar as the nonaxisymmetric
component, but results are given showing the additional effect of the
spiral arms. An axisymmetric potential is also considered, to compare
with the main features that arise in the nonaxisymmetric case.

This paper is organised as follows. In Section \ref{catalogo}, the
observational stellar sample is described. In Section \ref{sec:model},
the properties of the three-dimensional Galactic potential employed to
compute orbits are summarised. In Section \ref{sec:method} the new
method and its technique to analyse resonant trapping is introduced.
Section \ref{periodicas} gives a study of periodic orbits on the
Galactic plane obtained in a Galactic potential including only the bar
in its nonaxisymmetric components; this study gives the orbital support
on which the proposed method is based. Section \ref{zefec} shows the
important influence that resonant families on the Galactic plane can
have on stellar orbits at different $z$-distances from this plane,
which is fundamental for the application of the method to stars in
the Galactic halo. In Section \ref{fourier} this method is complemented
by presenting the spectral analysis of the Galactic orbits of the sample
stars and of the computed periodic orbits. In Section \ref{ejemplo} the
method is applied to the Kapteyn moving group in the Galactic halo.
Finally, Section \ref{concl} presents our conclusions.

%%%%%%%%%%%%%%%%%%%%%%%%%%%%%%%%%%%%%%%%%%%%%%%%%%%%%%%%%%%%%%%%%%%%%%%
\section{The Observational Sample of Stars}
\label{catalogo}

The sample used in this work consists of 1642 halo and disc stars from
the catalogues of $uvby$--$\beta$ photometry of high-velocity and
metal-poor stars observed at the National Astronomical Observatory,
San Pedro M\'artir, M\'exico (hereafter SPM) \citep{SN88,SPC93,SMM06,
SSC12} with 711, 553, 442, and 143 stars in each of these catalogues,
respectively. However, not all of these stars have complete kinematic
data, and some of these stars have been repeated in more than one
catalogue, which has been amended in the final data base as explained
in \cite{SSC12}. Also, about half of the stars in the first catalogue
\citep{SN88} have been observed at the Danish telescopes, La Silla,
Chile, (hereafter LS): 36.0\% of the stars in this first catalogue were
observed exclusively in the north at SPM, those with the note `N'
in their Table VIII; 50.1\% exclusively in the south at LS, those with
`S'; and 13.8\% at both observatories, denoted with `N/S'.

The coordinates and proper motions employed in this paper have been
taken primarily from: Hipparcos \citep{ESA97}, Tycho--2 \citep{H00},
and the revised NLTT \citep{SG03}. For a few stars other sources
have been used for the proper motions, such as the original NLTT
\citep{L79a}, the LHS \citep{L79b}, the Lowell Proper Motion Survey
\citep{Gic59,Gic61,Gic75}, and the PPM Star Catalogue \citep{RB91}.
In general the proper-motion errors range from 0.5--3.0~mas/yr per
component for the best sources, such as Hipparcos and Tycho--2, to as
large as $\ga 10$~mas/yr per component for the other sources.
The median errors of the proper motions are $\pm1.51$ mas/yr for
$\mu_{\alpha}$, and $\pm1.20$ mas/yr for $\mu_{\delta}$.

The distances were derived from the photometric absolute-magnitude
calibration of \citet{SBM04,SMM06} and \citet{SSC12},
or from parallaxes in SIMBAD for stars with errors less than 10\% or
for stars outside the colour range of the photometric calibration.  The
median error of these distances is $\pm6.8$ pc, or $\pm$9.6\% of the
distance.

The radial velocities have been selected from the literature, from a
number of sources, such as \citet{CLL94,BPF94,BF00,RN91,FS86,AB72}
and Nordstr\"om (private communication). SIMBAD has been consulted
to obtain the most precise and up-to-date values possible; about 20\%
of our radial velocities have been updated from our previous study
\citep{SSC12}. The errors of these radial velocities range from a few
tenths of a km s$^{-1}$ for the best sources (Carney et al. 1994;
Nordstr\"om, private communication), to $\approx 7$ km s$^{-1}$ for the
older sources.  The median error of the radial velocities is
$\pm1.45$ km s$^{-1}$.

%%%%%%%%%%%%%%%%%%%%%%%%%%%%%%%%%%%%%%%%%%%%%%%%%%%%%%%%%%%%%%%%%%%%%%
\section{The Galactic Model}
\label{sec:model}

In order to construct a comprehensive orbital study of stars in
`moving groups' in the Galactic disc and halo, a detailed,
observationally restricted, semi-analytic (i.e. with some of the
functions numerically solved), three-dimensional model of the Milky
Way galaxy, composed of axisymmetric and nonaxisymmetric potentials,
is employed \citep{PMA12}. In this model, some modifications are made
on the axisymmetric Galactic potential of \citet{AS91}. This
axisymmetric potential consists of three components: a disc, a
spherical bulge, and a massive spherical halo. The first modifications
to build the nonaxisymmetric Galactic potential consist in that all
the mass in the spherical bulge is employed to build the Galactic bar,
and a small fraction of the total mass of the disc is employed to
build the three-dimensional spiral arms. Therefore, only two
axisymmetric components are conserved in the final model, the slightly
diminished disc and the spherical halo. A boxy bar is employed to
represent the Galactic bar; its model is described in
\citet{PMM04}. The model for the three-dimensional spiral arms, called
 {\tt PERLAS}, is described in detail in \citet{PM03}. The
Galactic potential has been rescaled to the Sun's galactocentric
distance, $R_0$ = 8.3 kpc, and the local rotation velocity,
${\Theta}_0$ = 239 km s$^{-1}$, given by \citet{BRet11}.

A brief description of the observational/theoretical parameters
utilised to restrict the nonaxisymmetric components of the model is
presented next and summarised in Table \ref{tab:model} (further
details on the parameters of the model and restrictions were
introduced in \citet{PMA12}).

%%%%%%%%%%%%%%%%%%%%%%%%%%%%%%%%%%%%%%%%%%%%%%%%%%%%%%%%%%%%%%%%%%%%%%%
\subsection{The Galactic Bar}\label{barra}

In \citet{PMM04} three different models for the Galactic bar were
introduced, which approximate the density model of \citet{F98} of
COBE/DIRBE observations of the Galactic centre: a prolate, a triaxial
and a triaxial boxy bar. For the computations in this work, we have
adopted the triaxial boxy bar model, that fits better the Galactic
observations. Table \ref{tab:model} summarises the bar's structural
parameters, such as: the major semi-axis, scale lengths, and axial
ratios. The considered total mass for the bar, 
1.6 $\times$ 10$^{10}$ M$_{\odot}$, is within observational
estimations that lie in the range 1 -- 2 $\times$ 10$^{10}$ M$_{\odot}$
\citep[e.g.,][]{K92,Z94,DAH95,B95,SUet97,WS99}. The angular speed of
the bar and the present orientation of its major axis are still
controversial \citep[e.g.,][]{BGet91,WS99,BG02,BG05,MNQ07,G02}. We
consider $\phi$ = 20$^\circ$ for the present-day orientation of
the major axis of the Galactic bar; this is the angle between this 
major axis and the Sun-Galactic centre line. A long list of studies
have been presented in the literature that estimate the most important
dynamical parameter of the Galactic bar, its angular velocity,
${\Omega}_B$, \citep[and references therein]{G11}. \citet{G11}
concludes from his review that the most likely range in ${\Omega}_B$
is 50 -- 60 $\kmskpc$. For our computations, the value
${\Omega}_B$ = 55 $\kmskpc$ is employed, as listed in Table
\ref{tab:model}. This value is consistent with the recent estimate
of ${\Omega}_B$ given by \citet{Aet14}: ${\Omega}_B$ =
(1.89 $\pm$ 0.08)${\Omega}_o$, with ${\Omega}_o$ the local angular
velocity; in our potential this gives ${\Omega}_B$ = 54.4 $\pm$ 2.3
$\kmskpc$.

%%%%%%%%%%%%%%%%%%%%%%%%%%%%%%%%%%%%%%%%%%%%%%%%%%%%%%%%%%%%%%%%%%%%%%
\subsection{The Spiral Arms}\label{brazos}

The spiral-arm model employed in this work is the one called
{\tt PERLAS} \citep{PM03}; it is formed by a three-dimensional 
bisymmetric steady potential consisting of a superposition of
inhomogeneous oblate spheroids along a logarithmic spiral locus,
adjustable to better represent the available observations of the
Galactic spiral arms. This simulates the main Galactic spiral arms as
in \citet{BCHet05} and \citet{CHBet09}, based on the Spitzer/GLIMPSE
database. The position of the spiral arms at the present time is
taken as in figure 1 in \citet{PMA12}, i.e. the line where these arms
emerge at the inner Galactic region lags behind the major axis of the
bar making an angle of 40$^\circ$ with this axis.
The specific parameters of the spiral arms, such as angular
velocity, pitch angle, etc., are provided in Table \ref{tab:model} and
more specific details of the parameters are presented in \citet{PMA12}.
The spiral-arm mass is distributed as an exponential decline along the
arms, and their strength is related with their total mass, which is a
small fraction of the disc mass. To quantify the strength of the arms, 
the function $Q_T$ \citep{ST80,CS81} is computed. The maximum value of
$Q_T$ over the radial extent of the spiral arms, called $Q_s$, is a
measure of the strength of the spiral arms. In agreement with
\citet{BVSL05}, and considering that for Sbc galaxies, as for our
Galaxy, $Q_s$ is approximately less than 0.25, the corresponding range 
in total mass for the spiral arms, calculated from {\tt PERLAS} model, 
is obtained with $M_{\rm arms}/M_{\rm disc}$ = 0.04$\pm$0.01. 
We have taken the central value 3.9 $\times$ 10$^9$ M$_{\odot}$ for the
mass of the spiral arms. Finally, in the same manner as for the Galactic
bar, there are different methods to determine the spiral-arm angular
velocity; for a review see \citet{G11}. We use the mean value of
${\Omega}_S$ = 25 $\kmskpc$, provided by these methods, as
listed in Table \ref{tab:model}.

\begin{table*}
\begin{minipage}{90mm}
\caption{Parameters of the Nonaxisymmetric Galactic Components,
 Sun's Galactocentric Distance and Local Circular Rotation Speed}
\label{tab:model}
\begin{tabular}{lcr}
 \hline
 \hline

Parameter & Value & References\\
 \hline

\multicolumn{3}{c}{\it Triaxial Boxy Bar}\\
 \hline
Major Semi-Axis                     & 3.13 kpc            & 1 \\
Scale Lengths                       & 1.7, 0.64, 0.44 kpc & 1 \\
Axial Ratios                        & 0.64/1.7, 0.44/1.7  &   \\
Mass                                & 1.6 $\times$ 10$^{10}$ M$_{\odot}$     &   \\
Angle Between Major Axis            &                     &   \\
\,\,\, and the Sun-GC Line          & 20$^{\circ}$        & 2 \\
Pattern Speed ($\Omega_B$)          & 55 $\kmskpc$  & 3,7 \\
 \hline
\multicolumn{3}{c}{\it Spiral Arms}\\
 \hline
Pitch Angle ($i$)                   & 15.5$^{\circ}$ & 4 \\
Scale Length ($H_{\star}$)          & 3.9 kpc        & 5 \\
$M_{\rm arms}/M_{\rm disc}$         &  0.04         &   \\
Mass                                &  3.9 $\times$ 10$^9$ M$_{\odot}$           &  \\
Pattern Speed ($\Omega_S$)          & 25 $\kmskpc$     & 3 \\
 \hline
$R_0$                               & 8.3 kpc       & 6 \\
${\Theta}_0$                        & 239 km s$^{-1}$  & 6 \\
 \hline
\end{tabular}

{           (1)~\citet{F98};
            (2)~\citet{G02};
            (3)~\citet{G11};
            (4)~\citet{D00};
            (5)~\citet{BCHet05};
            (6)~\citet{BRet11}; 
            (7)~\citet{Aet14}. }

\end{minipage}
\end{table*}

%%%%%%%%%%%%%%%%%%%%%%%%%%%%%%%%%%%%%%%%%%%%%%%%%%%%%%%%%%%%%%%%%%%%%%
\section{A New Method to Analyse Resonant Trapping} 
\label{sec:method}

The orbits of the 1642 stars in the catalogues of Section \ref{catalogo}
are computed for different forms of the Galactic potential, such as 
(1) an axisymmetric case (the rescaled potential of \citet{AS91}), (2)
a nonaxisymmetric case that includes the disc and halo as the
axisymmetric components, plus the boxy bar, and (3) the nonaxisymmetric
case (2) with the addition of the spiral arms to its nonaxisymmetric
components. All the orbits are computed backward in time up to
10$^{10}$ yr. Some parameters are obtained in each orbit, such as the
orbital energy per unit mass, $E$, and the $z$-component of angular
momentum per unit mass, $h$ (the $z$ axis being perpendicular to the
Galactic plane); these two important parameters in our method are
obtained in all the employed forms of the Galactic potential, and are
computed in the Galactic $inertial$ frame. Other computed parameters
are the Jacobi constant per unit mass, $J$, (in the case of the
nonaxisymmetric potential including only the boxy bar in its
nonaxisymmetric components; case 2 above), the minimum and maximum
distances from the Galactic centre, the maximum distance from the
Galactic plane, and the orbital eccentricity.

\begin{figure}
  \includegraphics[width=0.95\hsize]{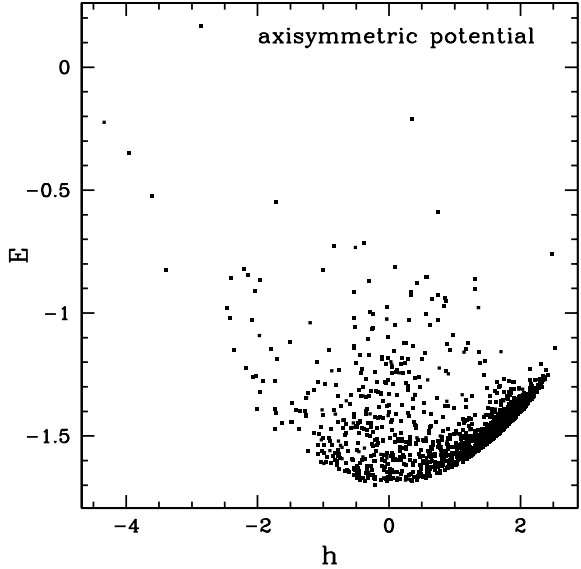}
  \caption{The Lindblad diagram of orbital energy per unit mass versus
  the $z$-component of orbital angular momentum per unit mass for the
  1642-star sample in the axisymmetric potential model of the
  Galaxy. The more bound (more negative orbital energies) and largest
  positive $z$-component orbital angular momenta orbits, represented
  by the darkest region to the lower-right part of the plot, includes
  mainly Galactic disc stars, while the remainder are mostly halo stars.
  The units for $h$ and $E$ are 10$^3$ km s$^{-1}$ kpc and
  10$^5$ km$^2$ s$^{-2}$, respectively.}
  \label{figura1}
\end{figure}

Figure \ref{figura1} shows results in the axisymmetric potential case.
In this potential the Galactic inertial frame is employed to compute
the orbits. For the 1642 stars, this figure shows the Lindblad diagram
$E$ versus $h$, both constants of motion in the axisymmetric potential.
Throughout this work, the units employed for $h$ are 
10$^3$ km s$^{-1}$ kpc, and 10$^5$ km$^2$ s$^{-2}$ for $E$ and $J$;
the unit of distance is kpc. As it can be noticed in Figure  
\ref{figura1}, in the axisymmetric case no structure is visible beyond
an approximate homogeneous disc (toward the lower right part of the
diagram), and an approximate homogeneous stellar halo (toward the
centre-left part of the diagram).

\begin{figure}
  \includegraphics[width=0.95\hsize]{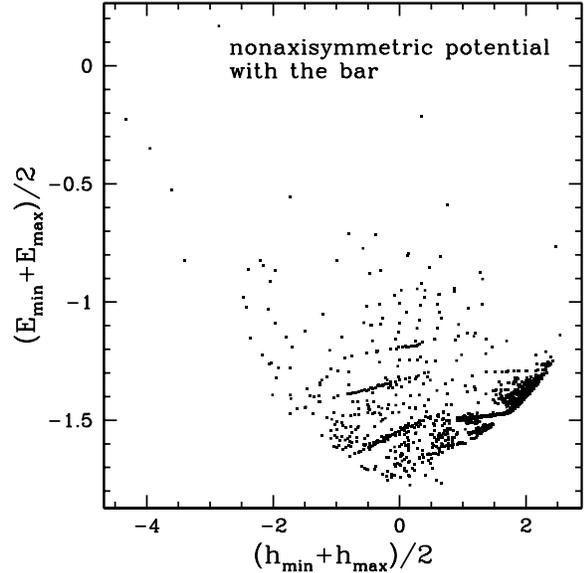}
  \caption{Similar to Figure \ref{figura1}, here for the Galactic
    potential model including only the boxy bar in its nonaxisymmetric
    components. In this case the diagram plots a `characteristic'
    orbital energy (in the Galactic inertial frame $E$ is not conserved
    along a given orbit) versus a `characteristic' orbital angular
    momentum ($h$ is not conserved either), both given by calculating
    the average of the maximum and minimum values of $E$ and $h$ along
    a given orbit. In this case, noticeable lines of dots
    (agglomerations of stars in this diagram) are formed in the halo
    and disc regions, induced by the presence of the bar.}
  \label{figura2}
\end{figure}

In a nonaxisymmetric potential, $E$ and $h$ computed in the Galactic
inertial frame are not conserved along the stellar orbit; thus, these
are not appropriate parameters to be compared in this case. After
several combinations of orbital parameters, it was found that the
particular combinations ($E_{\rm min}+E_{\rm max}$)/2 versus ($h_{\rm
  min}+h_{\rm max}$)/2 or ($E_{\rm min}+E_{\rm max}$)/2 versus $J$
gave interesting diagrams. $E_{\rm min}$, $E_{\rm max}$ and $h_{\rm
  min}$,$h_{\rm max}$ are the corresponding minimum and maximum values
of $E$ and $h$ along the stellar orbit. We call ($E_{\rm min}+E_{\rm
  max}$)/2 the `characteristic' orbital energy, and ($h_{\rm
  min}+h_{\rm max}$)/2 the `characteristic' orbital angular
momentum. In Figure \ref{figura2} the empirical diagram,
characteristic energy versus characteristic angular momentum, is
presented, employing a nonaxisymmetric potential that includes the
axisymmetric components, plus the central boxy bar (case (2) above).
This plot shows a noticeable difference with respect to the
axisymmetric case in Figure \ref{figura1}. In this case the
observational points gather in several bands that represent the
trapping regions produced by different resonances on the Galactic
plane, as it will be shown in Section \ref{periodicas}.  Although our
diagrams might seem similar to those of energy versus angular momentum
employed by \citet{HZ00} in their study of moving groups, the results
and origin of the formed moving groups are fundamentally different. In
the case of \citet{HZ00}, their method was applied to synthetic
satellites created from the beginning into an axisymmetric simplified
potential of the Galaxy. In this work, on the other hand, we produce a
comprehensive study by applying our method in a nonaxisymmetric
Galactic potential, where some moving groups might be shaped by
resonances generated by the Galactic bar.

\begin{figure}
  \includegraphics[width=0.95\hsize]{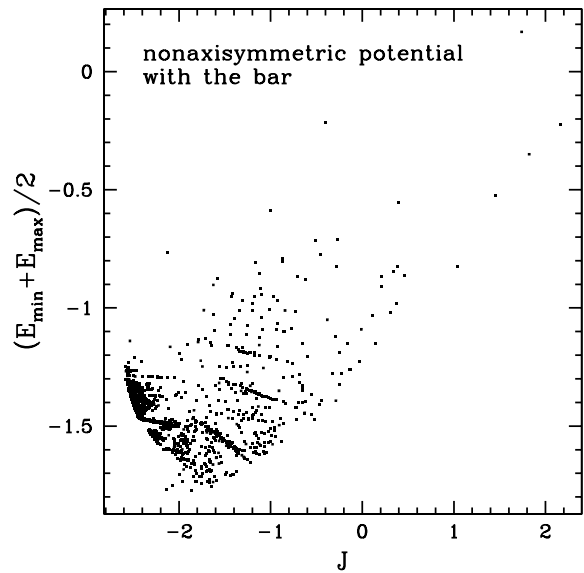}
  \caption{An alternate form to present the results in Figure
    \ref{figura2}; this diagram plots the characteristic orbital
    energy ($E_{\rm min}+E_{\rm max}$)/2 versus the orbital Jacobi
    constant, in units of 10$^5$ km$^2$ s$^{-2}$, computed in the
    reference frame of the bar. As in Figure \ref{figura2}, noticeable
    agglomerations of stars, induced by the presence of the bar, are
    formed in the halo and disc regions.}
  \label{figura3}
\end{figure}

In Figure \ref{figura3}, an alternate form to the results in Figure
\ref{figura2} is presented. This figure shows the characteristic
orbital energy ($E_{\rm min}+E_{\rm max}$)/2 versus the orbital Jacobi
constant $J$, which is conserved in the reference frame where the bar
is at rest. In the same manner as in Figure \ref{figura2}, dots
representing stars in the halo and disc agglomerate in noticeable lines
or groups, conserving the same structure obtained in that figure.
The crowded region of dots to the left in this diagram represents
mainly disc stars, while the region to the right contains mainly halo
stars.

\begin{figure}
  \includegraphics[width=0.95\hsize]{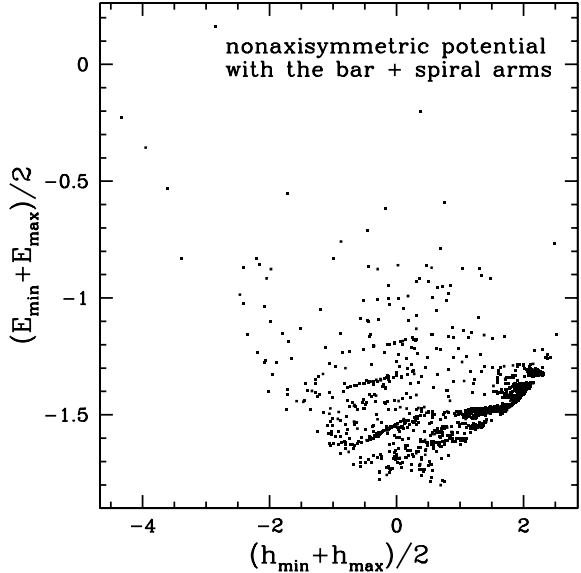}
  \caption{As in Figure \ref{figura2}, but now including the spiral
    arms into the nonaxisymmetric barred potential. Both figures are
    similar in the case of the halo stars; i.e., the spiral arms have
    little effect on halo stars, contrary to the case for disc stars
    shown by the most crowded region to the right of the diagram,
    where some differences can be noticed.}
  \label{figura4}
\end{figure}

Figure \ref{figura4} shows how the diagram in Figure \ref{figura2}
changes when the spiral arms are included in the nonaxisymmetric
potential. The difference between both diagrams is negligible for the
case of halo stars, i.e., the spiral arms have little effect on halo
stars, unlike the bar, where the effect on the formation of groups is
clear, most likely due to its much larger mass (approximately ten
times the mass of the spiral arms in the Milky Way case), and also
because the perigalacticons of these stars approach close or within
the bar region. However, the effect of the spiral arms is still of
importance on disc stars. In the following, results are given for the
Galactic potential including only the bar as a nonaxisymmetric
component; in a future study the effects of the spiral arms, in
connection with the results presented in the next sections, will be
analysed in detail.

\begin{figure}
  \includegraphics[width=0.95\hsize]{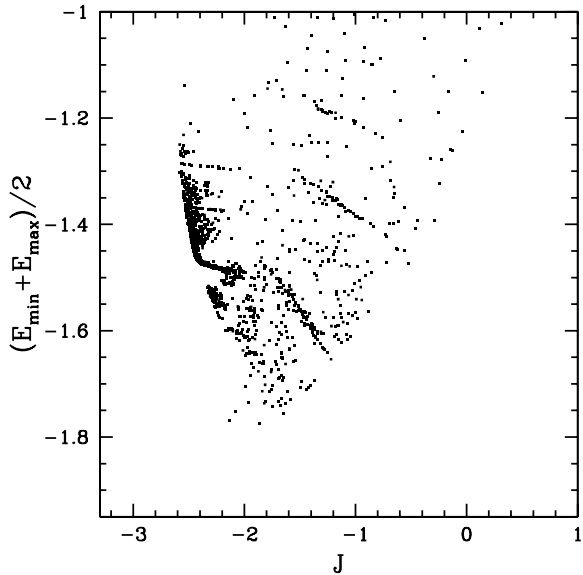}
  \caption{This figure shows a zoom of Figure \ref{figura3} to see in
   more detail the formation of the agglomerations in the disc and
   halo regions, and is at the same scale as the following Figure
   \ref{figura6}, for comparison.}
  \label{figura5}
\end{figure}

Figure \ref{figura5} is a zoom of Figure \ref{figura3}. In this plot
the agglomerations formed in the disc (the left-ward crowded region)
and in the halo (the remaining points) are illustrated in more detail.
This figure is employed in the next section.

The diagrams as in Figures \ref{figura2} and \ref{figura3} are the 
base of our method to delineate the agglomerations of points in the
disc and halo regions. These agglomerations are associated with
periodic orbits on the Galactic plane, as it is shown in the next
Section \ref{periodicas}. The next step is to investigate if known
moving groups in the disc and halo are contained in these regions.
In the following sections, some additional orbital properties are
presented related with these diagrams.

\section{Periodic Orbits on the Galactic Plane}\label{periodicas}

To connect with the diagrams shown in the previous section, several
families of periodic orbits $on$ $the$ $Galactic$ $plane$ have been
computed in the nonaxisymmetric Galactic potential, that includes the
axisymmetric components and the central boxy bar.  These families of 
periodic orbits are generated by the presence of the bar, and are 
defined in the $noninertial$ reference frame where the bar is at
rest.  In this frame, the x$'$,y$'$ plane is the Galactic plane, with
the x$'$ and y$'$ axes pointing along the major and minor axes of the
bar, respectively; the z$'$ axis coincides with the z axis in the
inertial reference frame.  In this inertial frame the Galactic and bar
rotations point in the clockwise direction.

The use of Poincar\'e-section diagrams has been useful to locate a 
starting periodic orbit in each family of periodic orbits, from which,
using a Newton--Raphson method \citep{PTVF92}, we have computed how the
orbit evolves as the value of the orbital Jacobi constant is varied.
For each computed orbit in a given family, the characteristic energy
($E_{\rm min}+E_{\rm max}$)/2, employed in the previous section, and
the orbital Jacobi constant $J$ are obtained; thus, the entire family
can be plotted in a diagram as in Figure \ref{figura3}.

\begin{figure}
  \includegraphics[width=0.95\hsize]{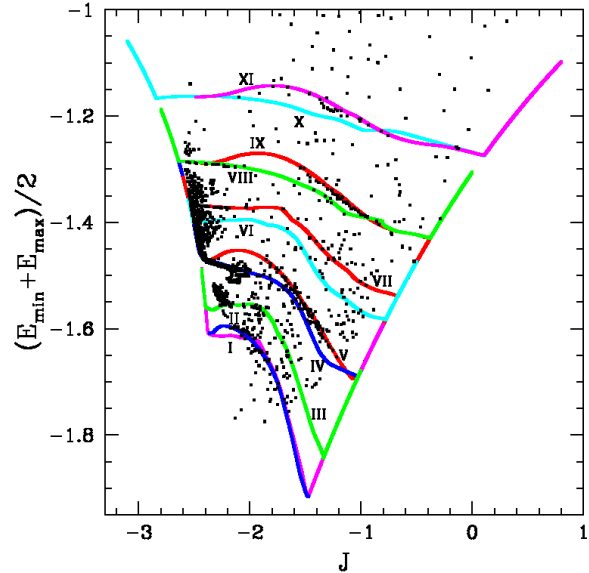}
  \caption{Characteristic energy versus $J$ for the nonaxisymmetric 
    barred Milky Way model, showing in coloured solid lines and with
    Roman numbers from I to XI, different families of periodic orbits
    $on$ $the$ $Galactic$ $plane$. These families are defined in the 
    noninertial reference frame where the bar is at rest. The black
    points are the observational points plotted in Figure
    \ref{figura5}. Notice that several families coincide with crowded
    lines of dots in Figure \ref{figura5}.} 
  \label{figura6}
\end{figure}

In Figure \ref{figura6} several families of periodic orbits, marked
with Roman numbers from I to XI, are plotted in a characteristic
energy vs $J$ diagram, along with the observational points in Figure
\ref{figura5}. These families are shown by coloured solid lines.
A given colour may represent more than one family. Notice that several
of these families of periodic orbits coincide with the crowded lines of
dots (groups of stars) induced by the bar (see Figure \ref{figura5}).
This occurs even in the halo region of the background observational
points, which is surprising, given that the periodic orbits lie on
the Galactic plane. This influence of resonances extending with
distance from the Galactic plane is analysed in Section \ref{zefec},
for specific families of periodic orbits in Figure \ref{figura6}.

\begin{figure}
  \includegraphics[width=0.95\hsize]{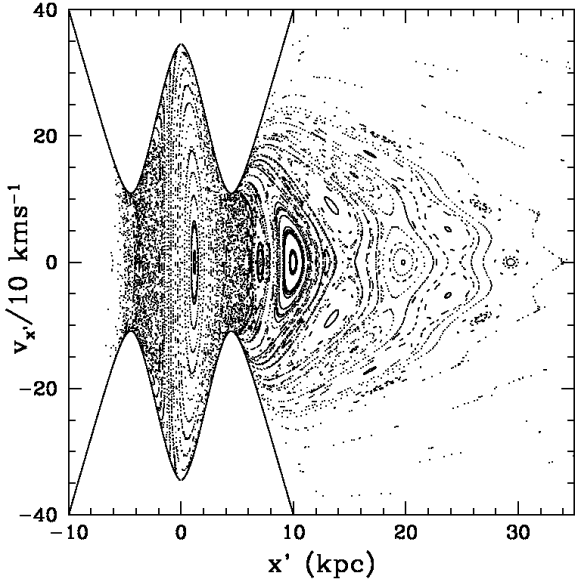}
  \caption{For stellar motion on the Galactic plane, this figure
  shows the Poincar\'e diagram in the plane $x',v_{x'}$ for
  $J$= $-$2.3 units.}
  \label{figura7}
\end{figure}

\begin{figure}
  \includegraphics[width=0.95\hsize]{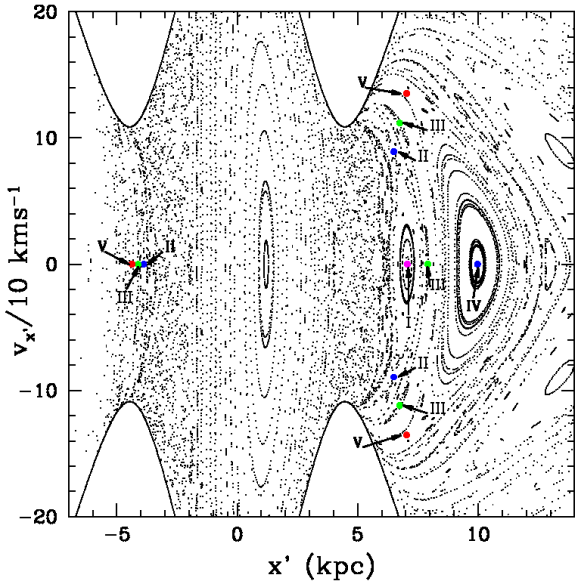}
  \caption{This is the left part of Poincar\'e diagram in Figure
  \ref{figura7}, showing the positions of families I to V in Figure
  \ref{figura6}. The colour of points for each family is the same as
  that employed in Figure \ref{figura6} for the corresponding curves.} 
  \label{figura8}
\end{figure}

\begin{figure}
  \includegraphics[width=0.95\hsize]{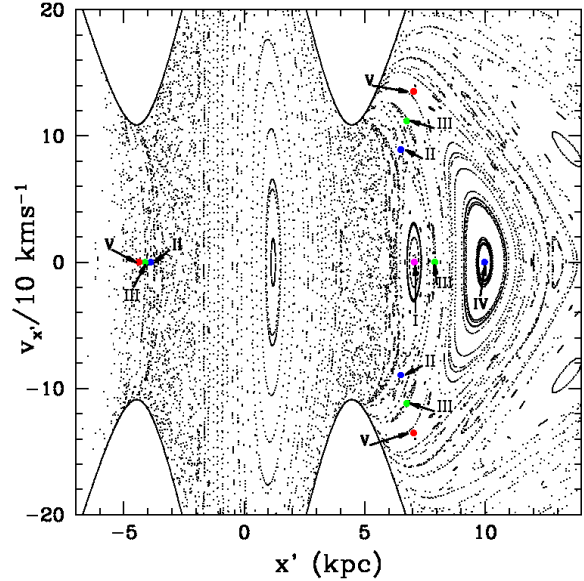}
  \caption{Here we show the positions in the Poincar\'e diagram for
  families VI to XI in Figure \ref{figura6}. Again, the colour of points
  corresponds to the colour of curves in Figure \ref{figura6}.}
  \label{figura9}
\end{figure}

For stellar motion on the Galactic plane analysed in the noninertial
reference frame, in Figure \ref{figura7} we show the Poincar\'e
diagram in the plane $x',v_{x'}$ for the value of the Jacobi constant
$J$= $-$2.3 units. Several resonant orbits are shown in this diagram.
In Figures \ref{figura8} and \ref{figura9} we mark in this Poincar\'e
diagram the families of periodic orbits I to XI considered in Figure
\ref{figura6}. Notice that for this particular value of $J$ the
corresponding orbit in a given family can be stable or unstable, i.e.
encircled or not by invariant curves. 

\begin{figure}
  \includegraphics[width=0.95\hsize]{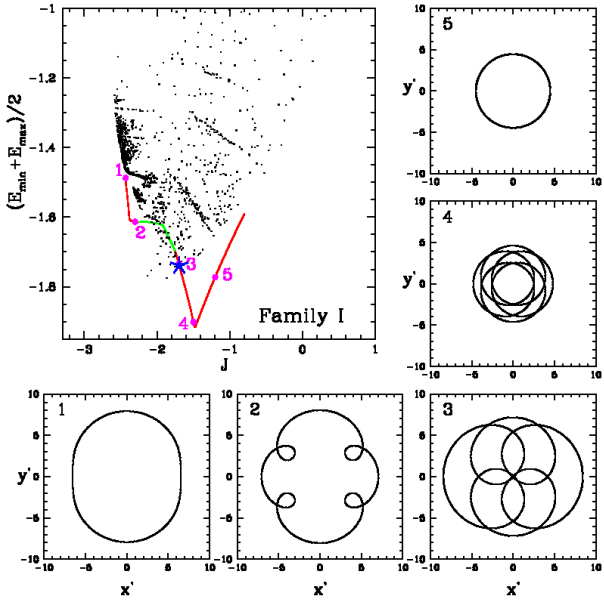}
  \caption{This is family `I' in Fig. \ref{figura6}. The stable and
    unstable parts along the curve are shown with red and green
    colours, respectively. Five periodic orbits along the family curve
    are illustrated, numbered from 1 to 5 on the curve and in their
    frames. The orbits are plotted in the noninertial reference frame
    where the bar is at rest. The x$'$ and y$'$ axes on the Galactic
    plane point along the major and minor axes of the bar,
    respectively. The blue star on the family curve shows the
    approximate position where there is a change in orbital rotation
    as seen in the $inertial$ Galactic frame: from retrograde to
    prograde going from right to left. In the inertial frame the
    Galactic and bar rotations point in the clockwise
    direction. Distances are in kpc.}
  \label{fig:famper1}
\end{figure}

\begin{figure}
  \includegraphics[width=0.95\hsize]{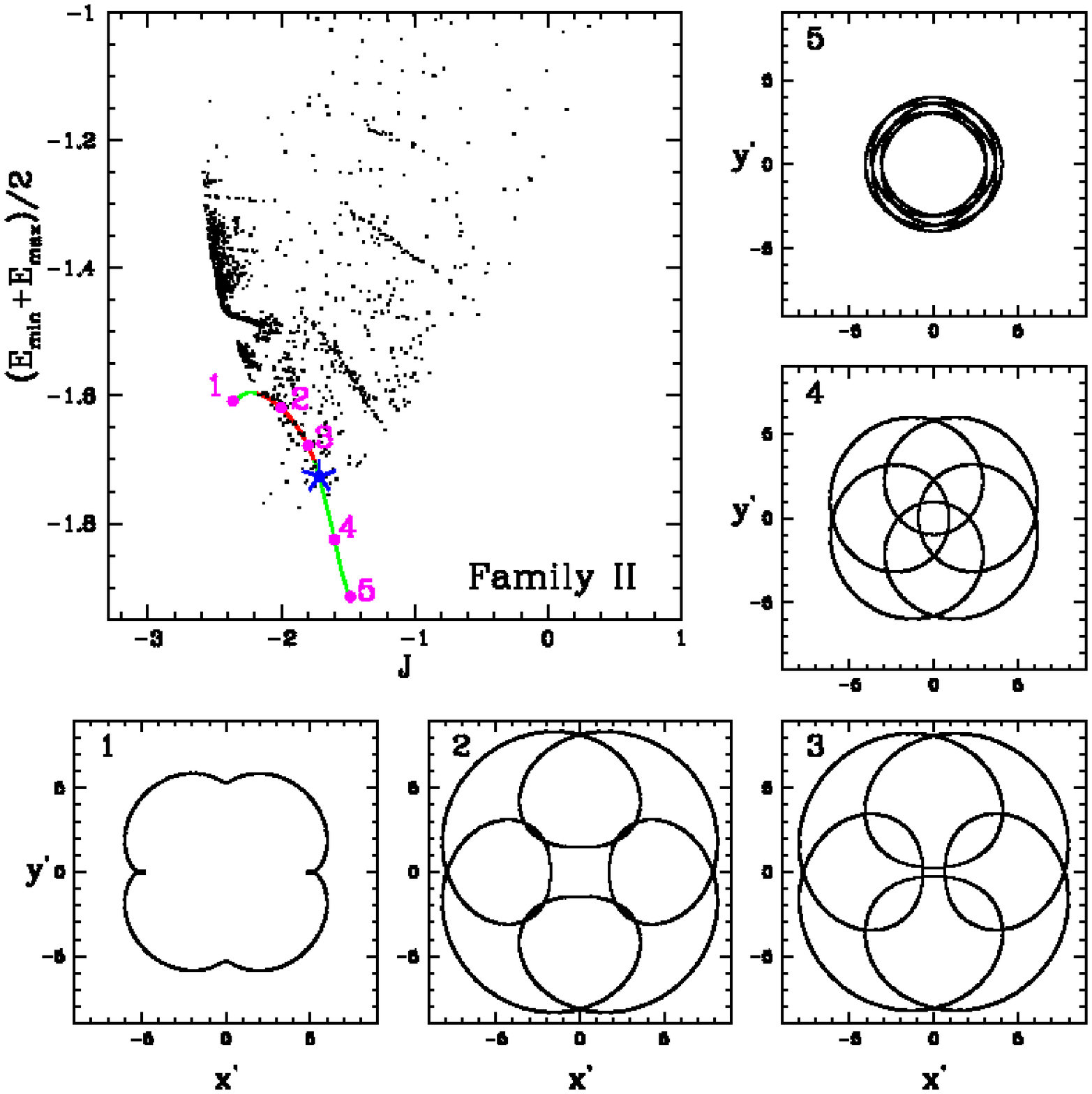}
  \caption{The same as Figure \ref{fig:famper1}, here for the family
    labelled as `II' in Figure \ref{figura6}.}
  \label{fig:famper2}
\end{figure}

\begin{figure}
  \includegraphics[width=0.95\hsize]{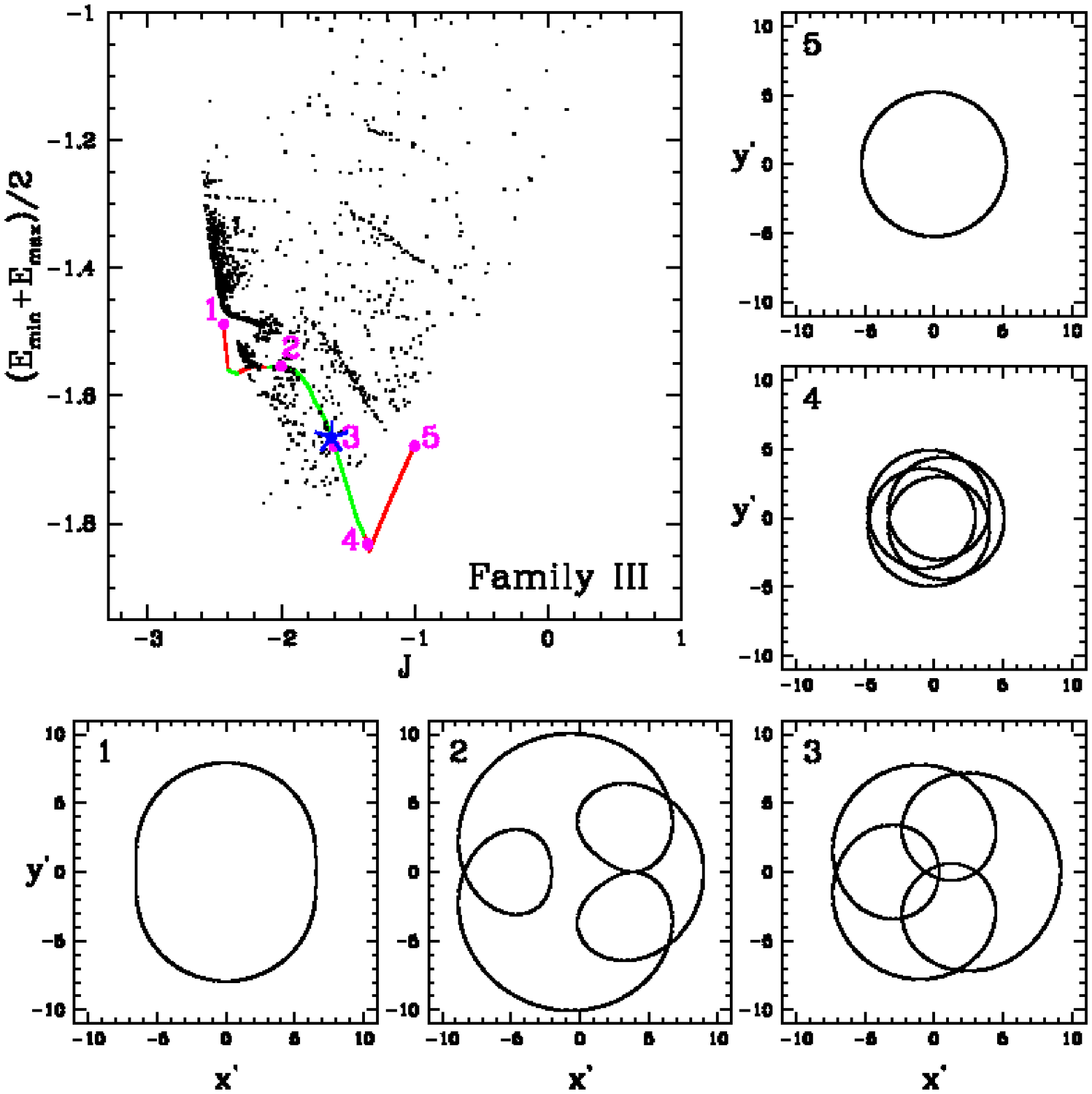}
  \caption{The same as Figure \ref{fig:famper1}, here for the family
    labelled as `III' in Figure \ref{figura6}.}
  \label{fig:famper3}
\end{figure}

\begin{figure}
  \includegraphics[width=0.95\hsize]{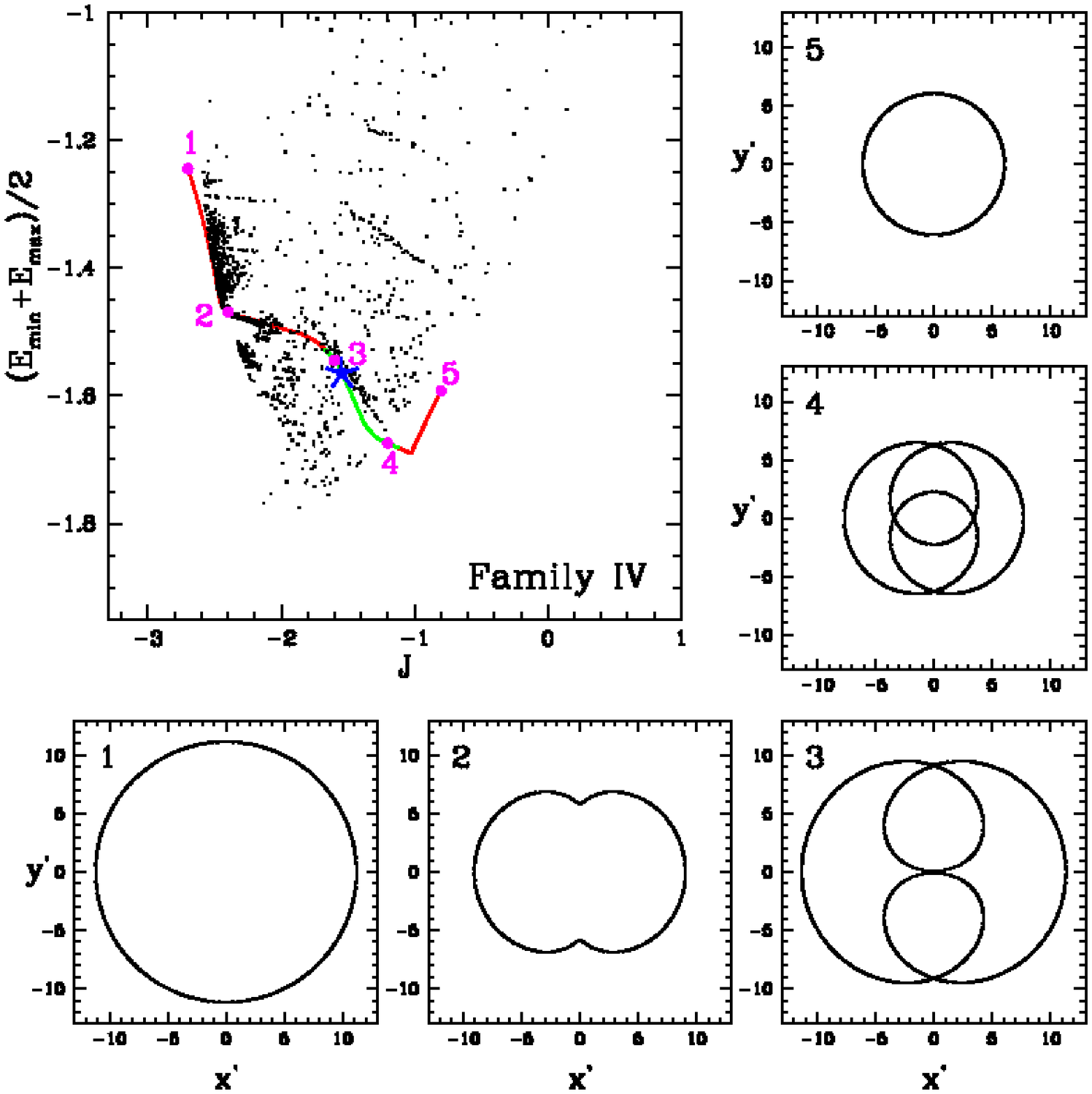}
  \caption{The same as Figure \ref{fig:famper1}, here for the family
    labelled as `IV' in Figure \ref{figura6}.}
  \label{fig:famper4}
\end{figure}

\begin{figure}
  \includegraphics[width=0.95\hsize]{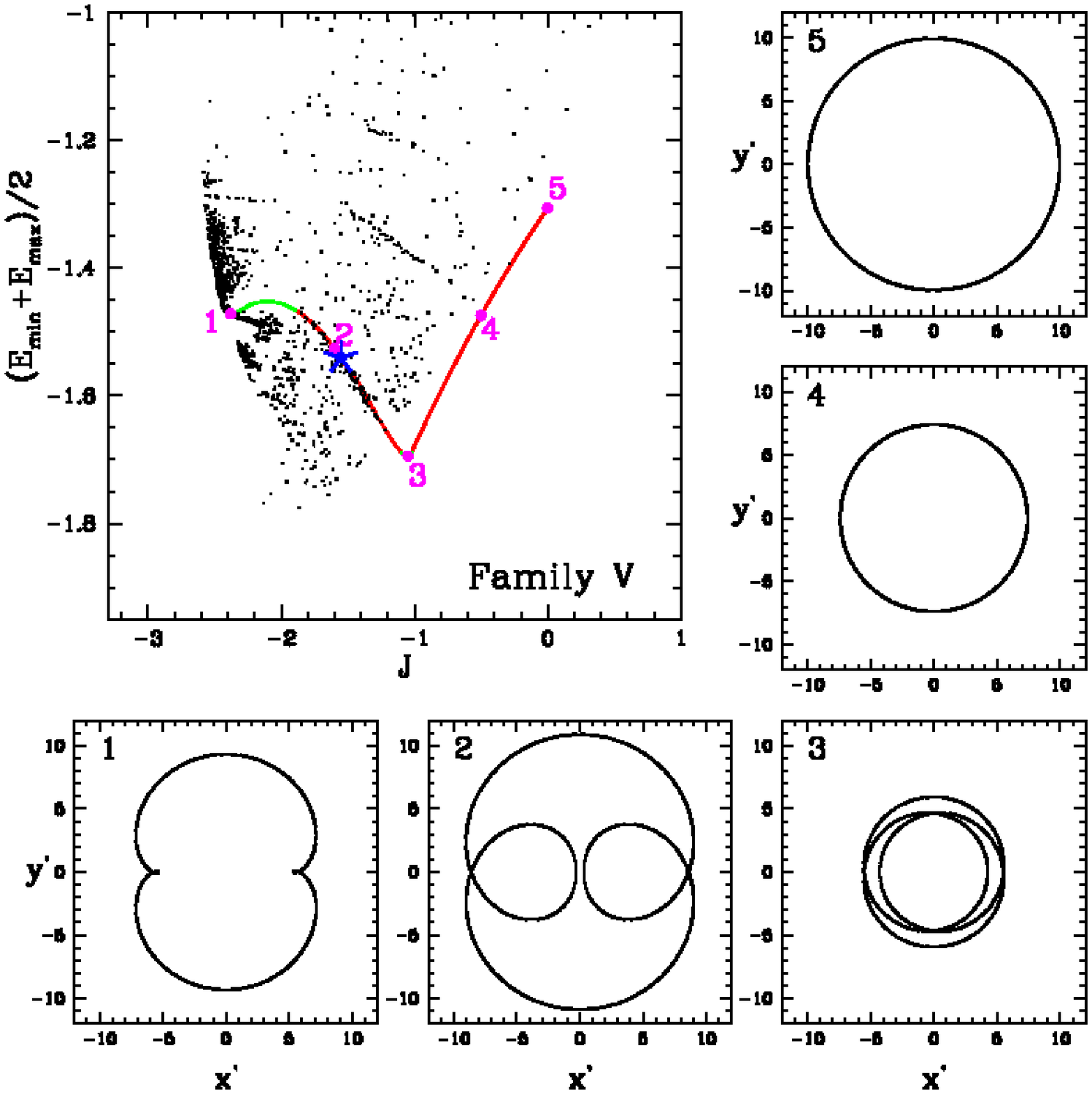}
  \caption{The same as Figure \ref{fig:famper1}, here for the family
    labelled as `V' in Figure \ref{figura6}.}
  \label{fig:famper5}
\end{figure}

\begin{figure}
  \includegraphics[width=0.95\hsize]{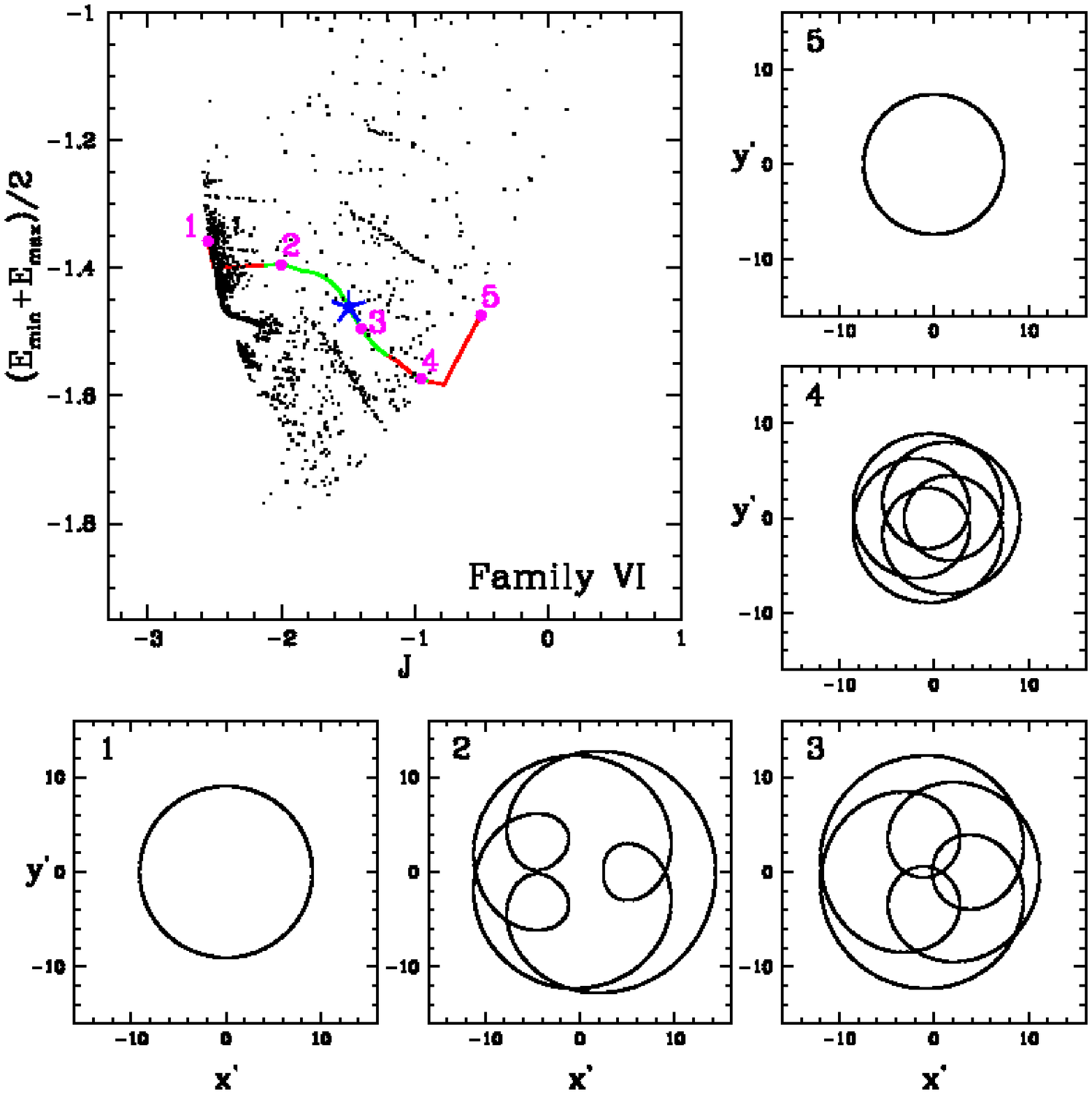}
  \caption{The same as Figure \ref{fig:famper1}, here for the family
    labelled as `VI' in Figure \ref{figura6}.}
  \label{fig:famper6}
\end{figure}

\begin{figure}
  \includegraphics[width=0.95\hsize]{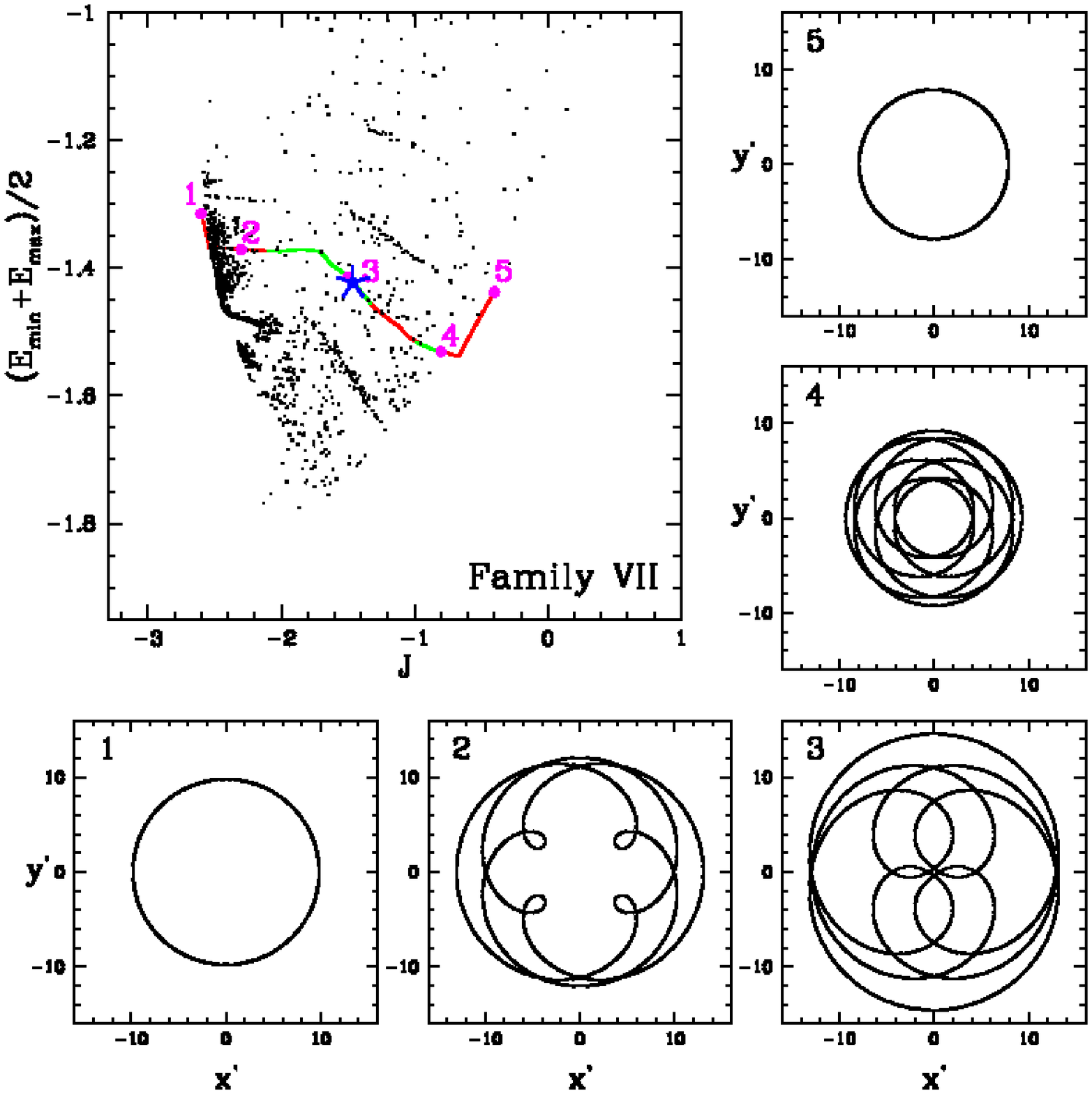}
  \caption{The same as Figure \ref{fig:famper1}, here for the family
    labelled as `VII' in Figure \ref{figura6}.}
  \label{fig:famper7}
\end{figure}

\begin{figure}
  \includegraphics[width=0.95\hsize]{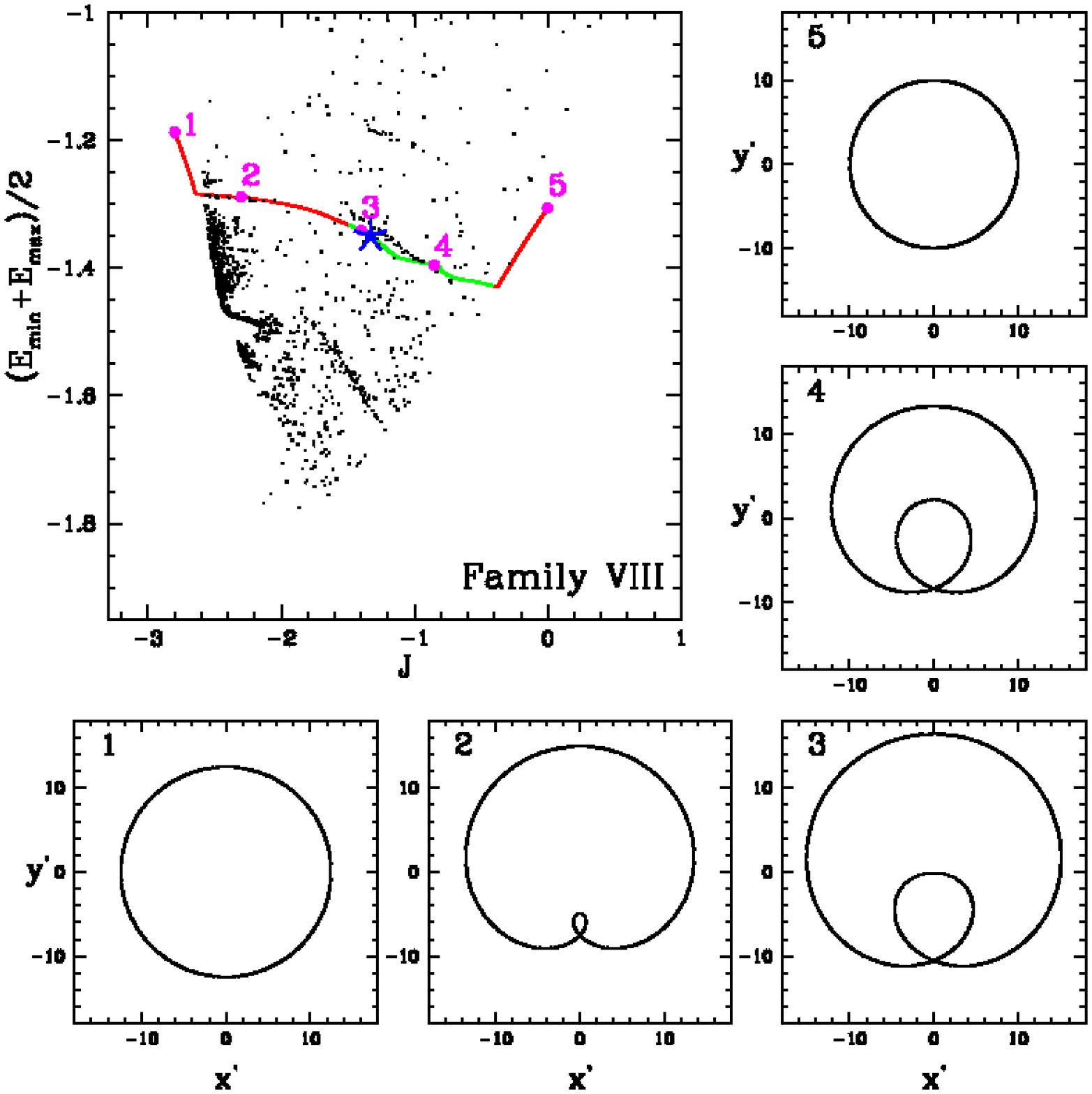}
  \caption{The same as Figure \ref{fig:famper1}, here for the family
    labelled as `VIII' in Figure \ref{figura6}.}
  \label{fig:famper8}
\end{figure}

\begin{figure}
  \includegraphics[width=0.95\hsize]{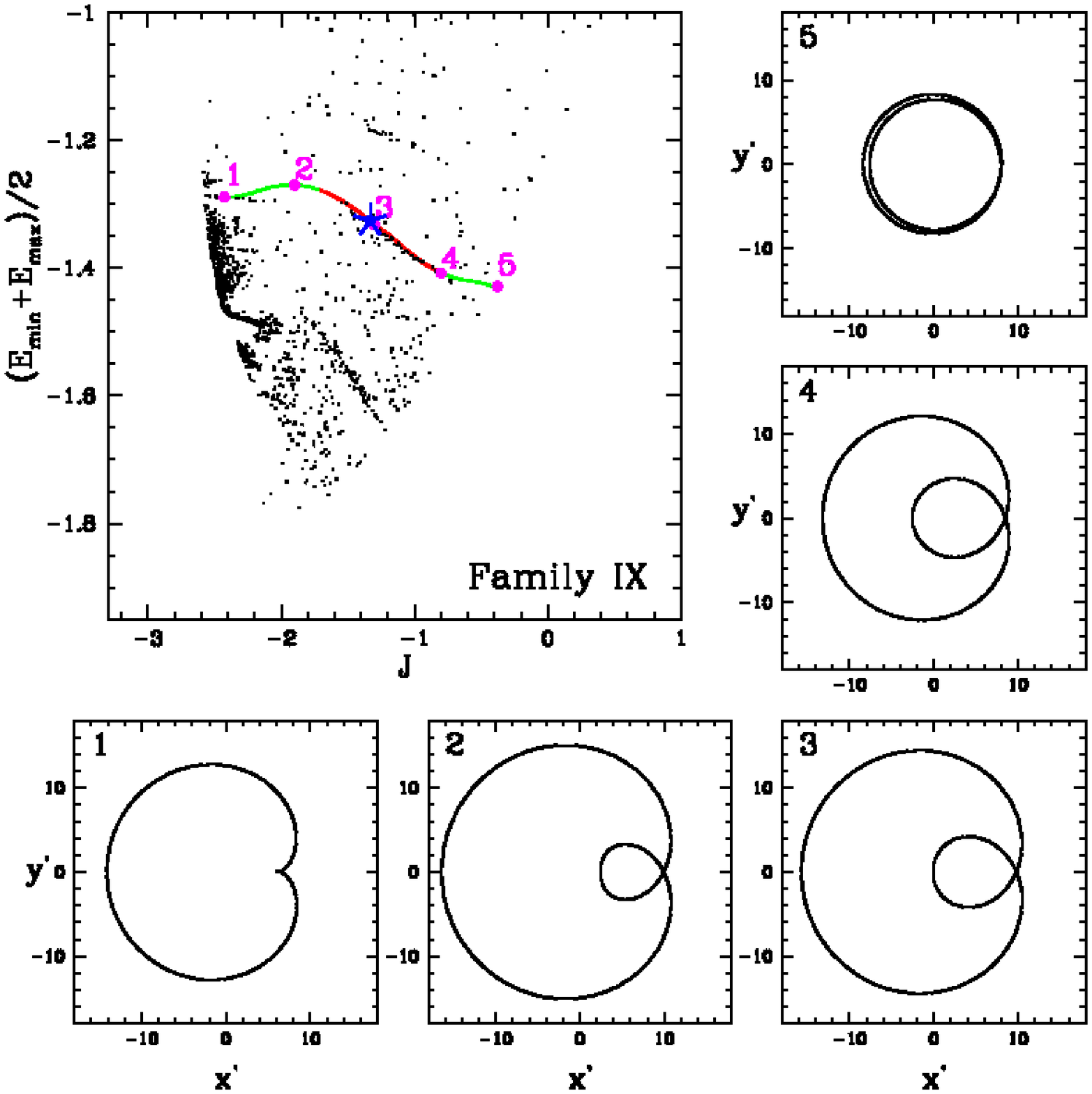}
  \caption{The same as Figure \ref{fig:famper1}, here for the family
    labelled as `IX' in Figure \ref{figura6}.}
  \label{fig:famper9}
\end{figure}

\begin{figure}
  \includegraphics[width=0.95\hsize]{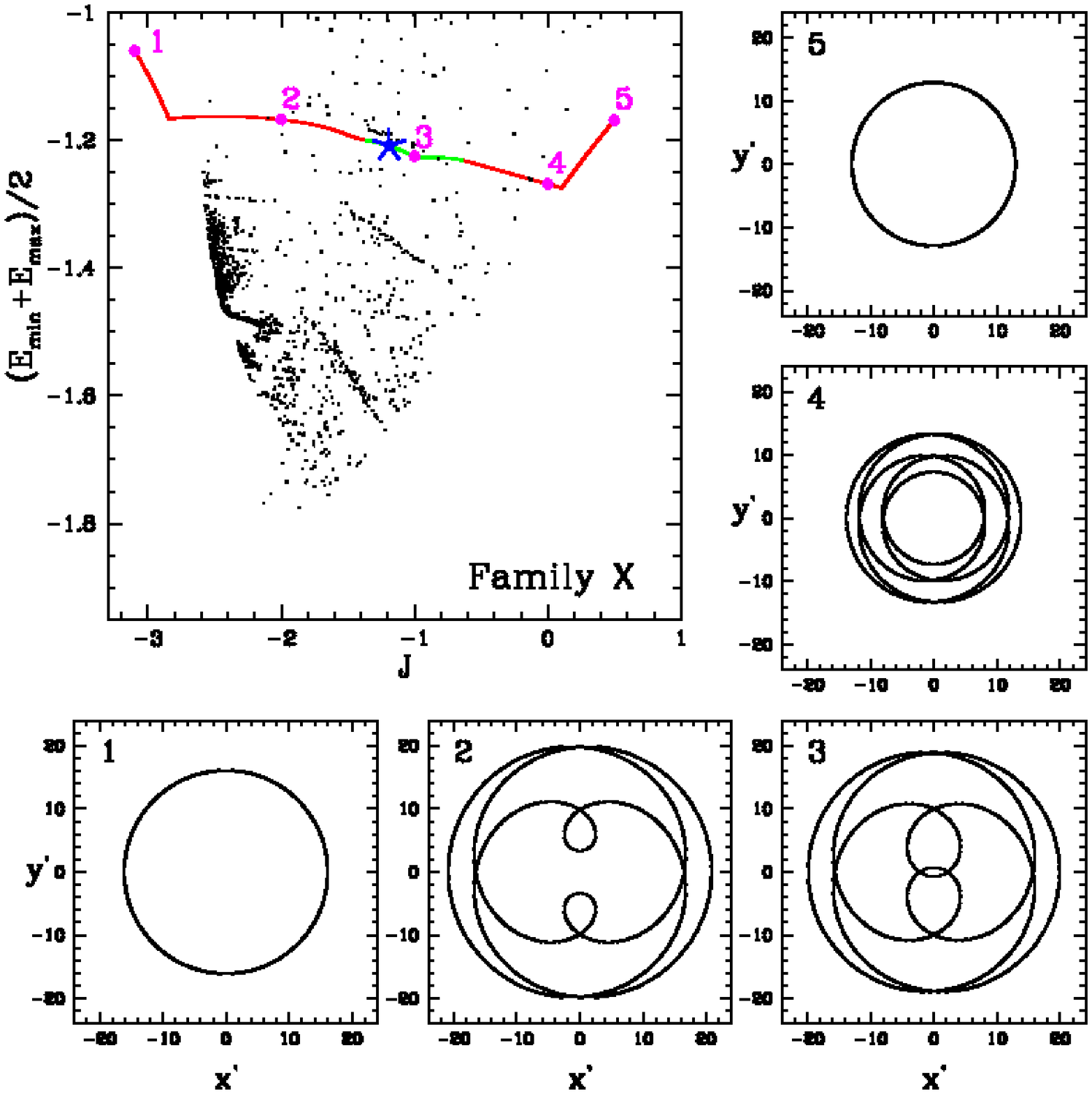}
  \caption{The same as Figure \ref{fig:famper1}, here for the family
    labelled as `X' in Figure \ref{figura6}.}
  \label{fig:famper10}
\end{figure}

\begin{figure}
  \includegraphics[width=0.95\hsize]{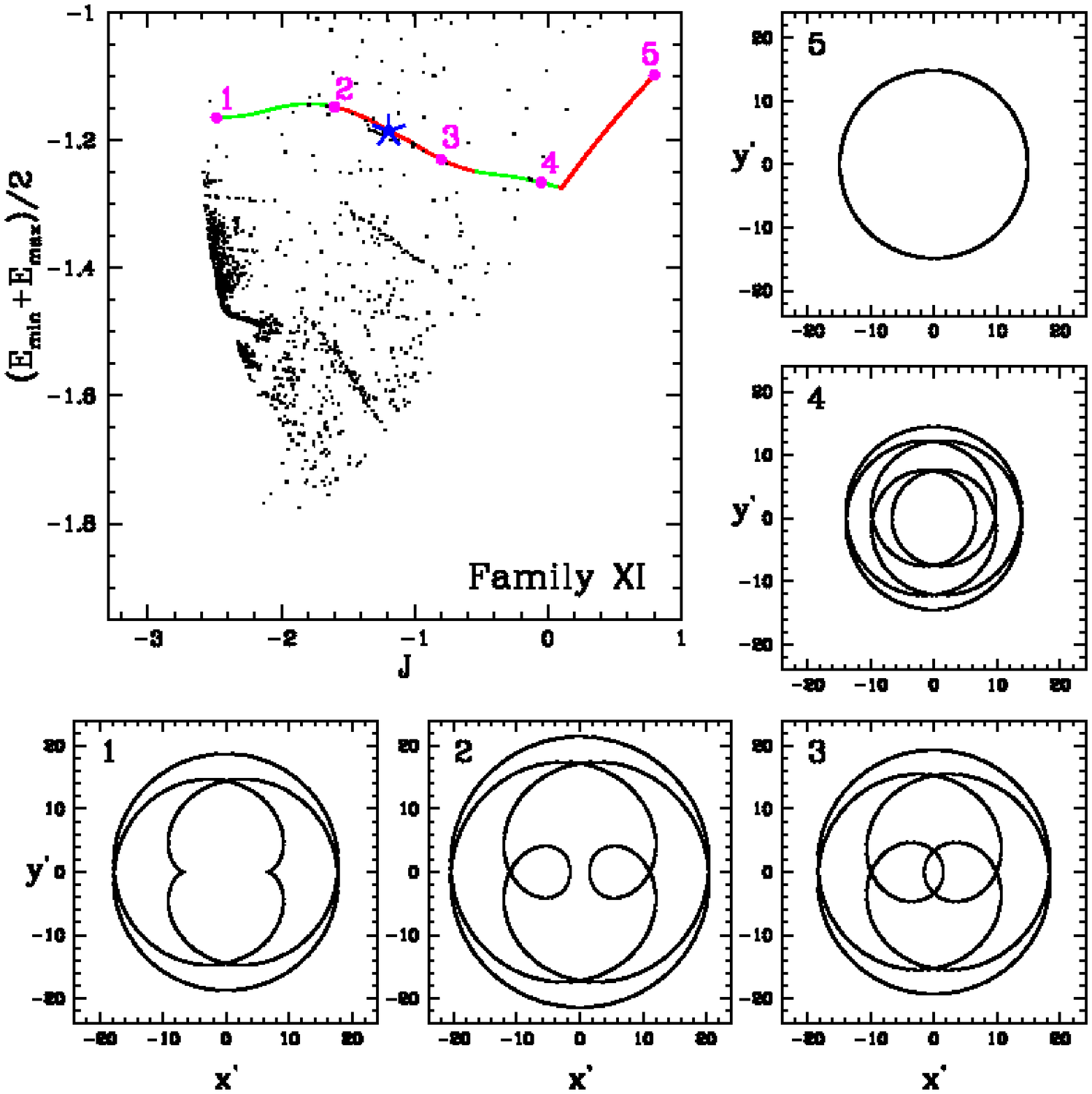}
  \caption{The same as Figure \ref{fig:famper1}, here for the family
    labelled as `XI' in Figure \ref{figura6}.}
  \label{fig:famper11}
\end{figure}

Figures \ref{fig:famper1} to \ref{fig:famper11} show each individual
family of periodic orbits given in Figure \ref{figura6}.  In these
figures the curve of the given family is plotted in two colours, red
and green, which show the stable and unstable parts in each curve,
respectively. The stability or instability of the computed periodic
orbits in each family was analysed following the method given by
\citet{HE65}. The agglomerations of points around a given family curve
may occur in its stable parts. Also, in these figures five positions
are taken in each family curve, labelled from 1 to 5, and their
corresponding periodic orbits are shown in the five small frames,
including the position number; the scale is the same in these frames,
with distances in kpc. These orbits are plotted in the noninertial
reference frame where the bar is at rest.

As Figure \ref{figura6} shows, all the family curves merge on both the
left and right boundaries of the region where the points distribute.
Figures \ref{fig:famper1} to \ref{fig:famper11} show that all the
details of the periodic orbits tend to disappear at these boundaries,
the orbits becoming almost circular. Due to the presence of the bar,
this approximation to circular orbits becomes strictly true as the
orbital characteristic energy increases.

With respect to the $inertial$ Galactic frame, the sense of orbital
rotation (of the corresponding orbit in this frame, which in general
will not be periodic) in these left and right boundaries in Figure
\ref{figura6} is as follows: in the right boundary, i.e., higher
values of $J$, the orbital rotation is $retrograde$; in the left
boundary, toward the region where the disc component distributes, this
rotation is $prograde$. In Figures \ref{fig:famper1} to
\ref{fig:famper11} a blue starred point on each curve marks the
approximate position where the sense of orbital rotation changes in
the $inertial$ frame, from retrograde to prograde and vice versa. This
position corresponds approximately to the periodic orbit in the family
that crosses the Galactic centre (nearly localised by one of the five
sample periodic orbits in the cases of families
I,III,IV,V,VII,VIII,IX; see corresponding figures); this orbit will
not have strictly $h$ = 0, as $h$ is not conserved.

An additional property of a given family of periodic orbits, is that a
point on its curve in a diagram as in Figure \ref{figura6}, i.e.,
characteristic energy versus $J$, is not necessarily associated with a
single periodic orbit. If a periodic orbit is not symmetric with
respect to one of the axes of the bar, as in families III,VI,VIII,IX,
the reflection of the orbit on this axis, which generates a different
orbit in the given x$'$,y$'$ frame, will give the same point in the
diagram as the original orbit; i.e., the characteristic energy and $J$
are the same in both orbits. The existence of the reflected orbit is
allowed due to the symmetries of the potential, i.e., the sense of
both the x$'$,y$'$ axes can be changed simultaneously, recovering the
original situation. Thus, both types of orbit coexist in the given
x$'$,y$'$ frame.

\section{Influence of Resonances on z-Distance}\label{zefec}

In Figure \ref{figura6} some families of periodic orbits on the
Galactic plane coincide with agglomerations of points in Figure
\ref{figura5}. These agglomerations occur in the disc and halo regions;
thus, the influence of these resonant families does not appear to be
restricted necessarily to a region close to the Galactic plane. This
influence can extend deep into the halo region.

\begin{figure}
  \includegraphics[width=0.95\hsize]{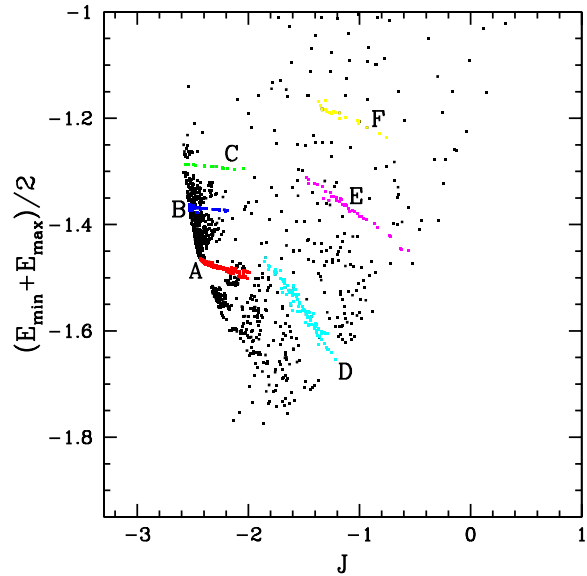}
  \caption{Six regions A,B,C,D,E,F of observational points from Figure 
   \ref{figura5}, shown in different colours.  Regions A,B,C are
   associated respectively with families IV,VII,VIII in Figure
   \ref{figura6} (see Figures \ref{fig:famper4}, \ref{fig:famper7},
   \ref{fig:famper8}). Regions D,E,F are associated with stable parts 
   of families V,IX,XI in Figure \ref{figura6}, respectively (see
   Figures \ref{fig:famper5}, \ref{fig:famper9}, \ref{fig:famper11}).}
  \label{figura21}
\end{figure}

To show this result, in Figure \ref{figura21} six agglomerations or
regions, A,B,C,D,E,F in different colours, have been considered from
the observational points in Figure \ref{figura5}. Regions A,B,C are
associated respectively with families IV,VII,VIII in Figure
\ref{figura6} (see Figures \ref{fig:famper4}, \ref{fig:famper7},
\ref{fig:famper8}). Regions D,E,F are associated respectively with
families V,IX,XI in Figure \ref{figura6}, which have stable parts in
these regions (see Figures \ref{fig:famper5}, \ref{fig:famper9},
\ref{fig:famper11}), but probably not with the corresponding
neighbouring families IV,VIII,X, which have unstable parts in these
regions (see Figures \ref{fig:famper4}, \ref{fig:famper8},
\ref{fig:famper10}).

\begin{figure}
  \includegraphics[width=0.95\hsize]{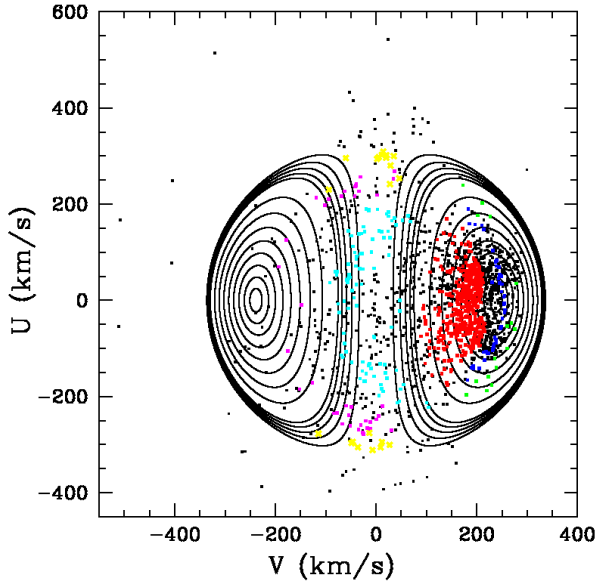}
  \caption{The 1642-star sample in the U,V velocity plane. These
  velocities are with respect to the inertial Galactic frame; U $<$ 0
  toward the Galactic centre. Regions A,B,C,D,E,F in Figure
  \ref{figura21} are shown with their corresponding colours. Some
  curves with constant orbital eccentricity in a Bottlinger diagram
  are also shown. From the inner to outer curves in the prograde and
  retrograde parts, the values of the eccentricity are 0.1,0.2,0.3,
  ...,0.9,0.92,0.94,0.96,0.98 .}
  \label{figura22}
\end{figure}

Figure \ref{figura22} shows the distribution of the 1642-star sample
on the plane of velocities U,V with respect to the inertial Galactic
frame, U being negative toward the Galactic centre, and V positive in
the direction of Galactic rotation. The points
corresponding to stars in the regions A,B,C,D,E,F in Figure
\ref{figura21} are shown with their colours in that figure,
along with a Bottlinger diagram showing some curves of constant
orbital eccentricity, obtained with the approximation of an inverse 
square law for the radial force \citep{TW53}. Regions D,E,F distribute
in the prograde-retrograde transition middle zone, with high orbital
eccentricities. 

\begin{figure}
  \includegraphics[width=0.95\hsize]{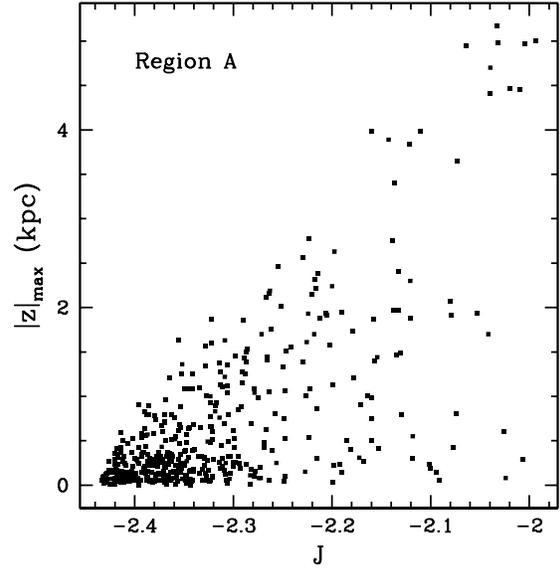}
  \caption{Maximum orbital distance from the Galactic plane for the
   stars in the region A, shown in Figure \ref{figura21}, versus their
   corresponding $J$.}
  \label{figura23}
\end{figure}

\begin{figure}
  \includegraphics[width=0.95\hsize]{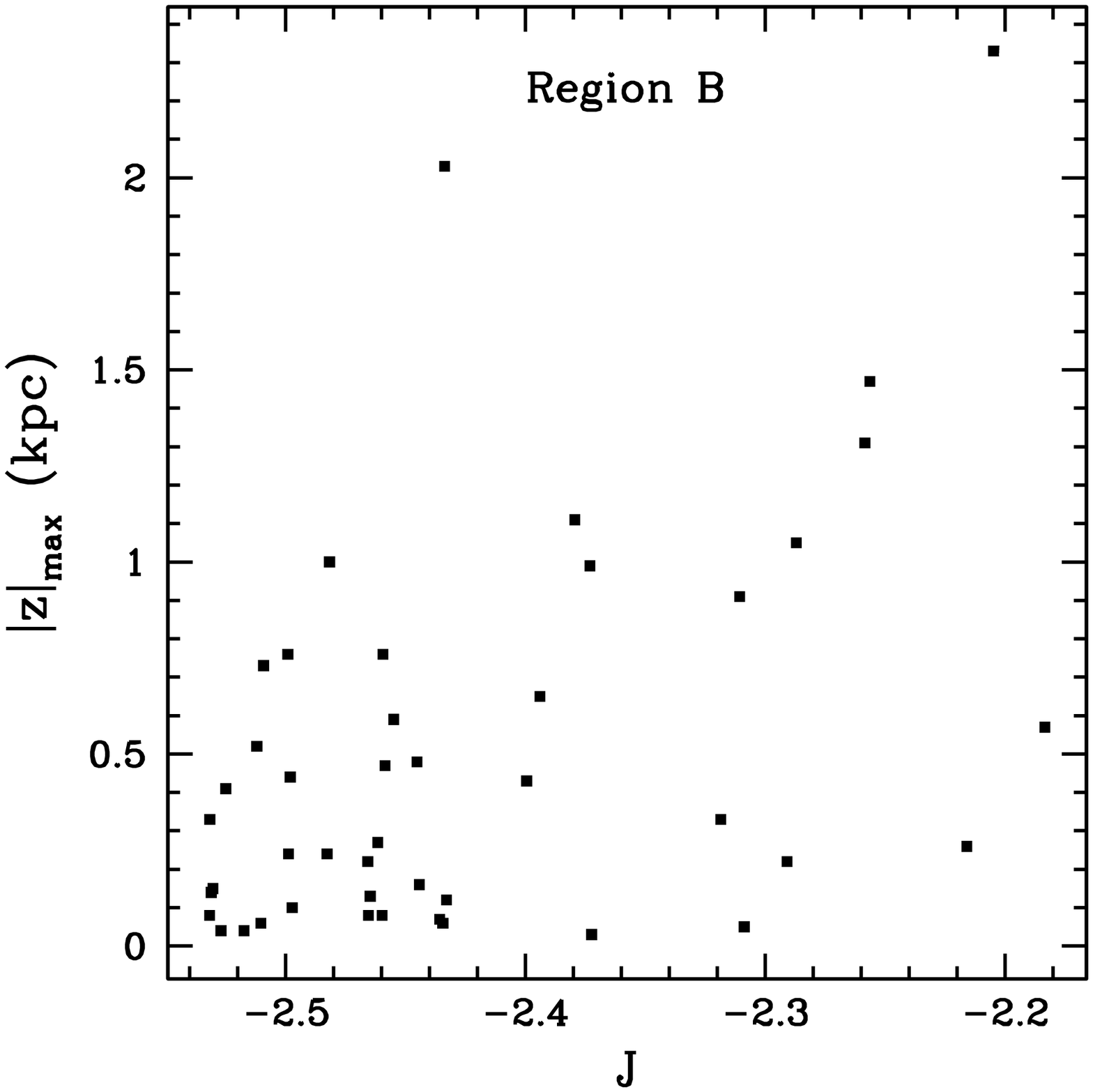}
  \caption{Same as in Figure \ref{figura23}, here for the region B of
   Figure \ref{figura21}.}
  \label{figura24}
\end{figure}

\begin{figure}
  \includegraphics[width=0.95\hsize]{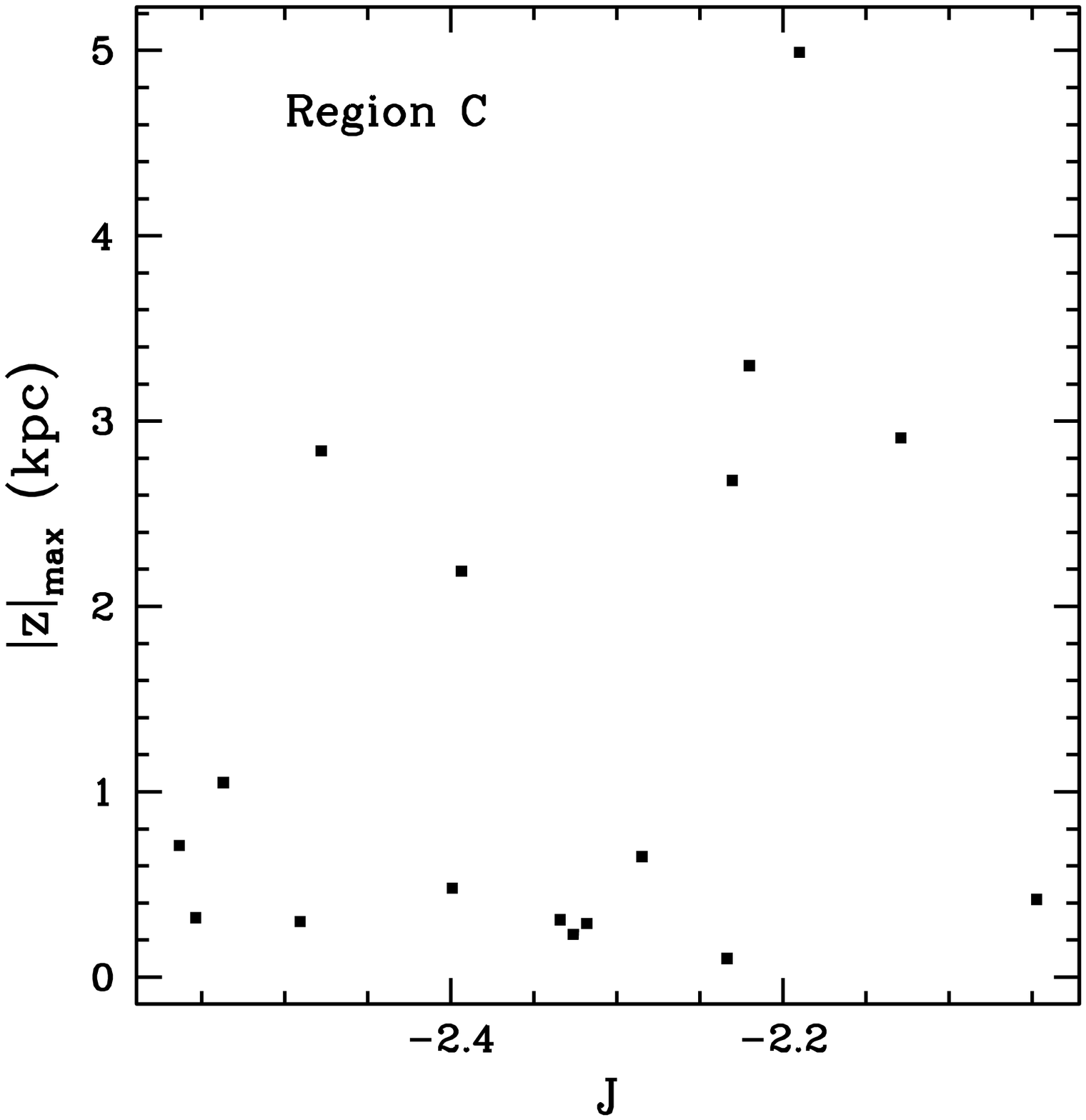}
  \caption{Same as in Figure \ref{figura23}, here for the region C of
   Figure \ref{figura21}.}
  \label{figura25}
\end{figure}

\begin{figure}
  \includegraphics[width=0.95\hsize]{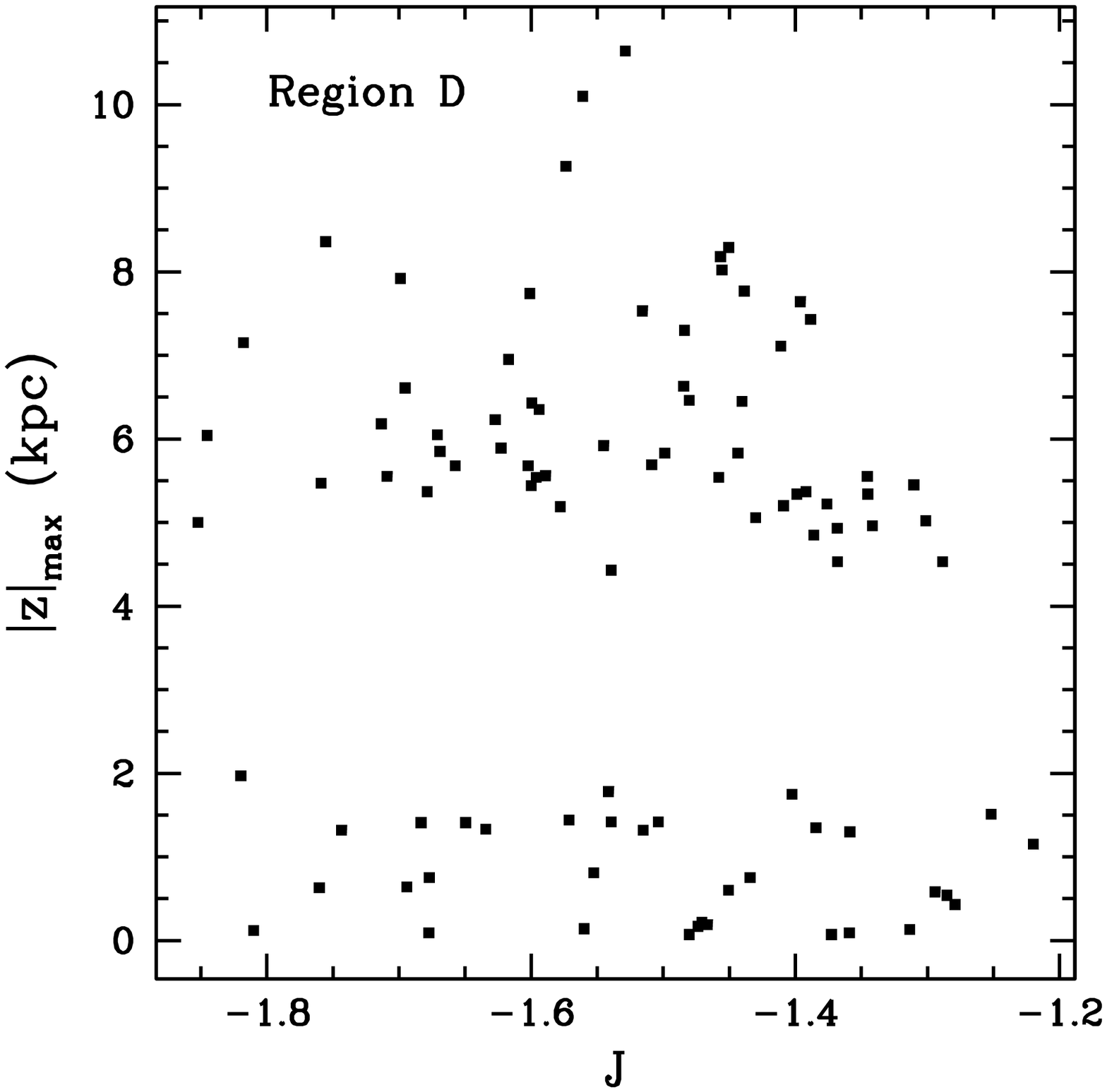}
  \caption{Same as in Figure \ref{figura23}, here for the region D of
   Figure \ref{figura21}.}
  \label{figura26}
\end{figure}

\begin{figure}
  \includegraphics[width=0.95\hsize]{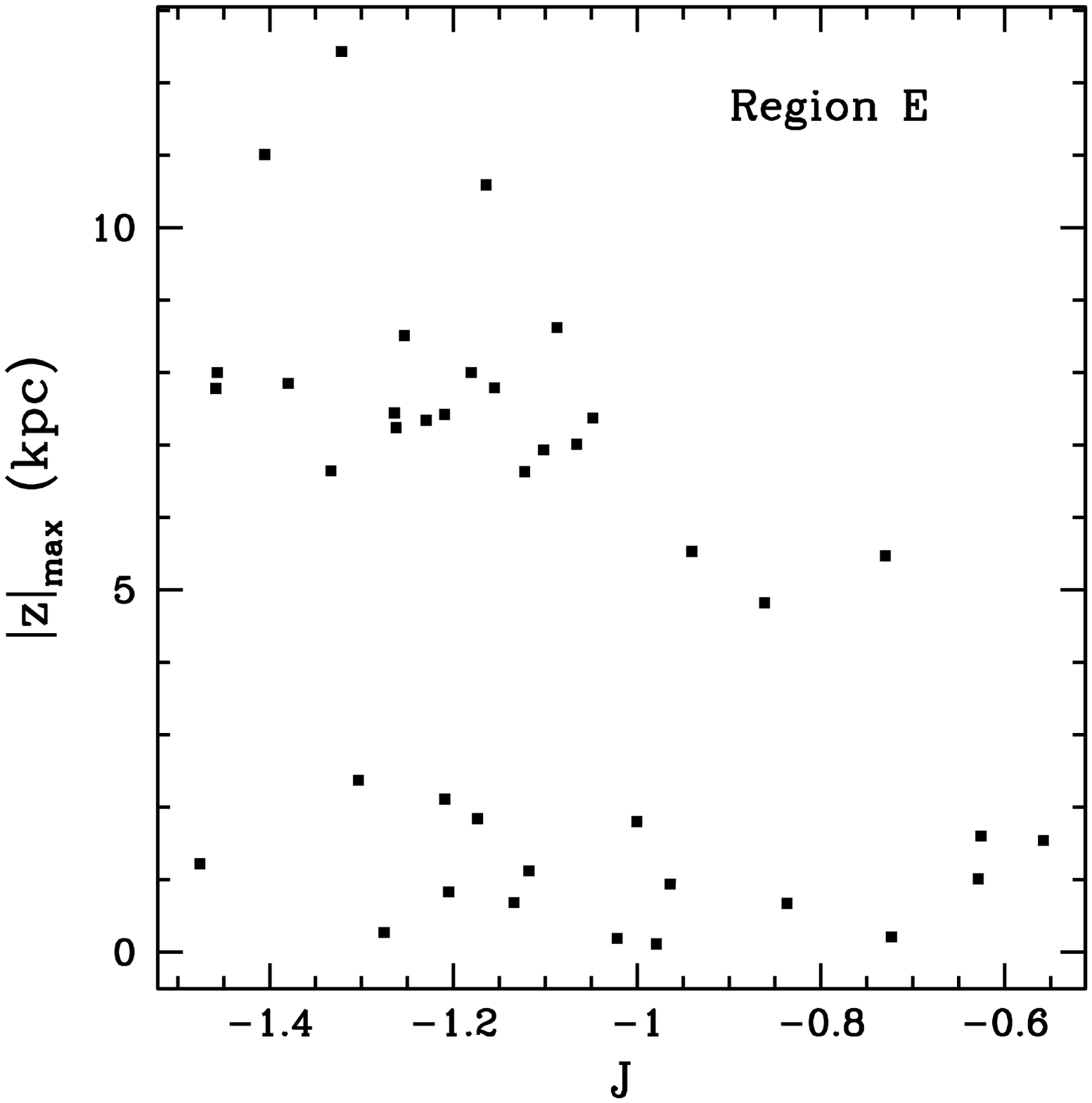}
  \caption{Same as in Figure \ref{figura23}, here for the region E of
   Figure \ref{figura21}.}
  \label{figura27}
\end{figure}

\begin{figure}
  \includegraphics[width=0.95\hsize]{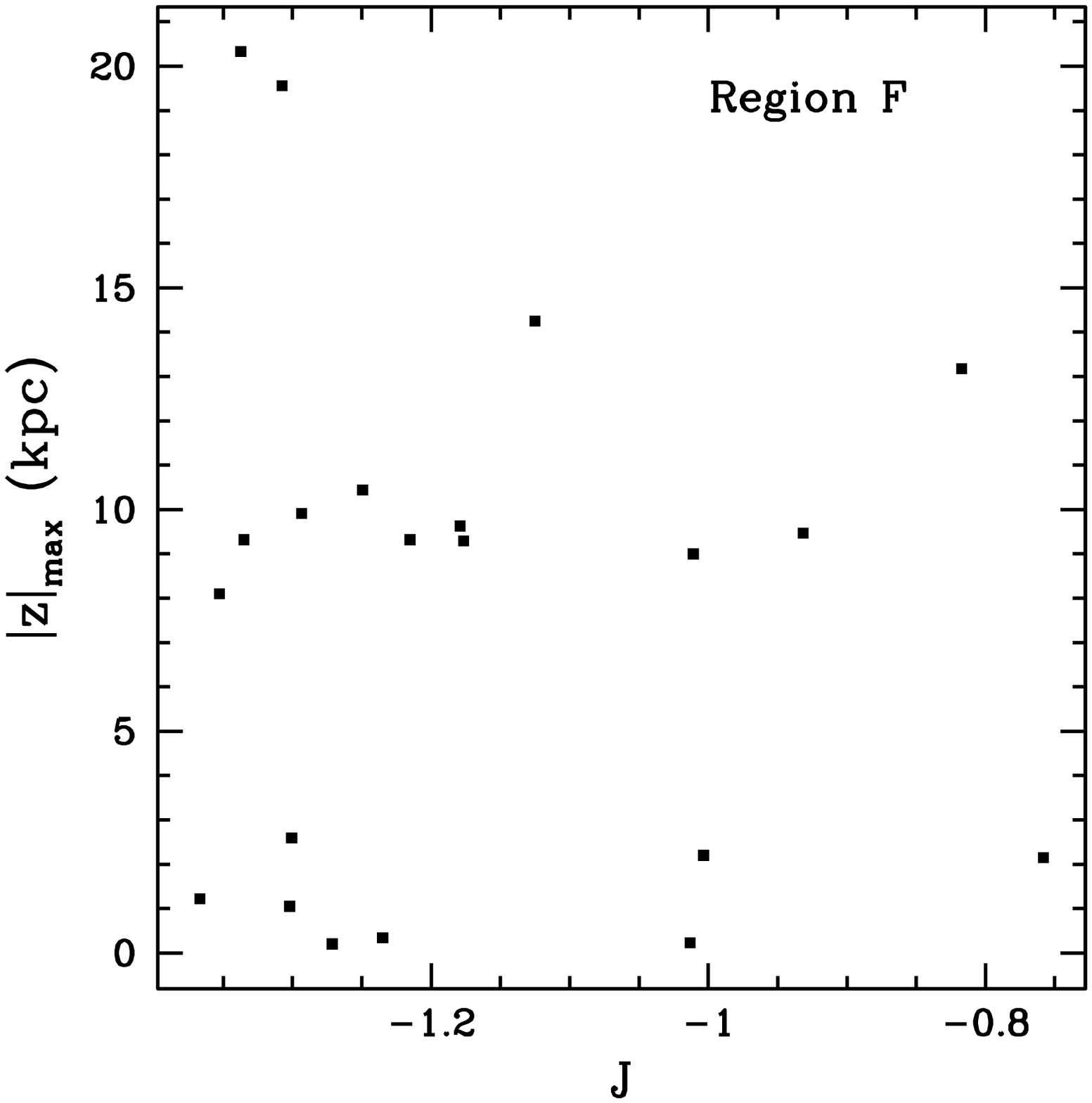}
  \caption{Same as in Figure \ref{figura23}, here for the region F of
   Figure \ref{figura21}.}
  \label{figura28}
\end{figure}

For every point (star) in each of the regions A,B,C,D,E,F, the maximum 
orbital distance from the Galactic plane, $|z|_{\rm max}$, was
computed, and corresponding diagrams $|z|_{\rm max}$ versus $J$ were
made for each region. These diagrams are shown in Figures \ref{figura23}
to \ref{figura28}. The $z$-distance can reach high values, especially
in the regions D,E,F for the halo component.

\begin{figure}
  \includegraphics[width=0.95\hsize]{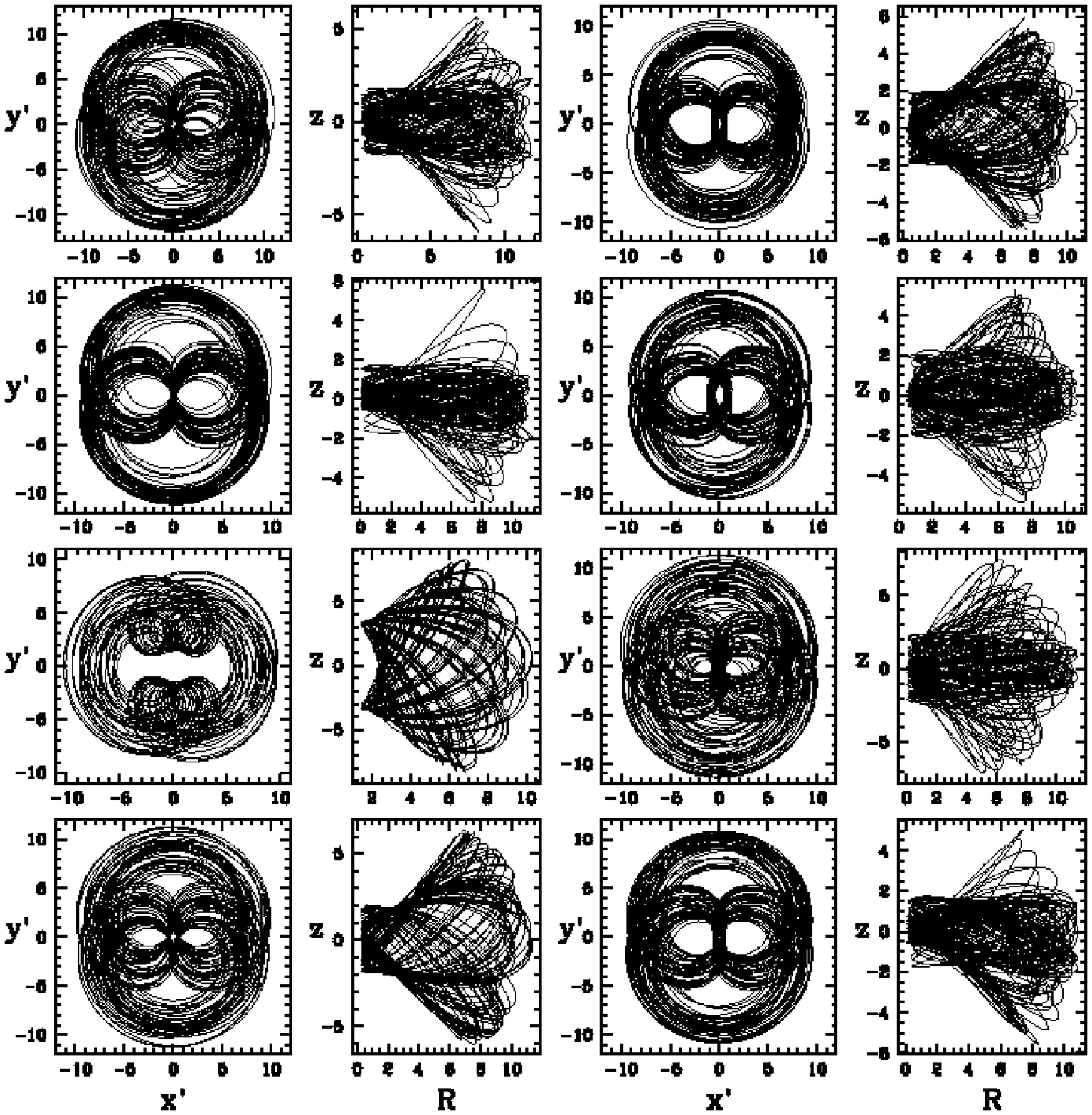}
  \caption{For a sample of stars reaching high $z$-distances and
   lying in region D in Figure \ref{figura21}, this figure shows 
   eight pairs of frames, four pairs are arranged in the first and
   second columns and the other four in the third and fourth
   columns. In each pair, the frame at the left gives the orbital
   projection on the Galactic plane x$'$,y$'$, in the reference
   frame where the bar is at rest, and the frame at the right gives 
   the meridional orbit in the plane $R,z$.}
  \label{figura29}
\end{figure}

\begin{figure}
  \includegraphics[width=0.95\hsize]{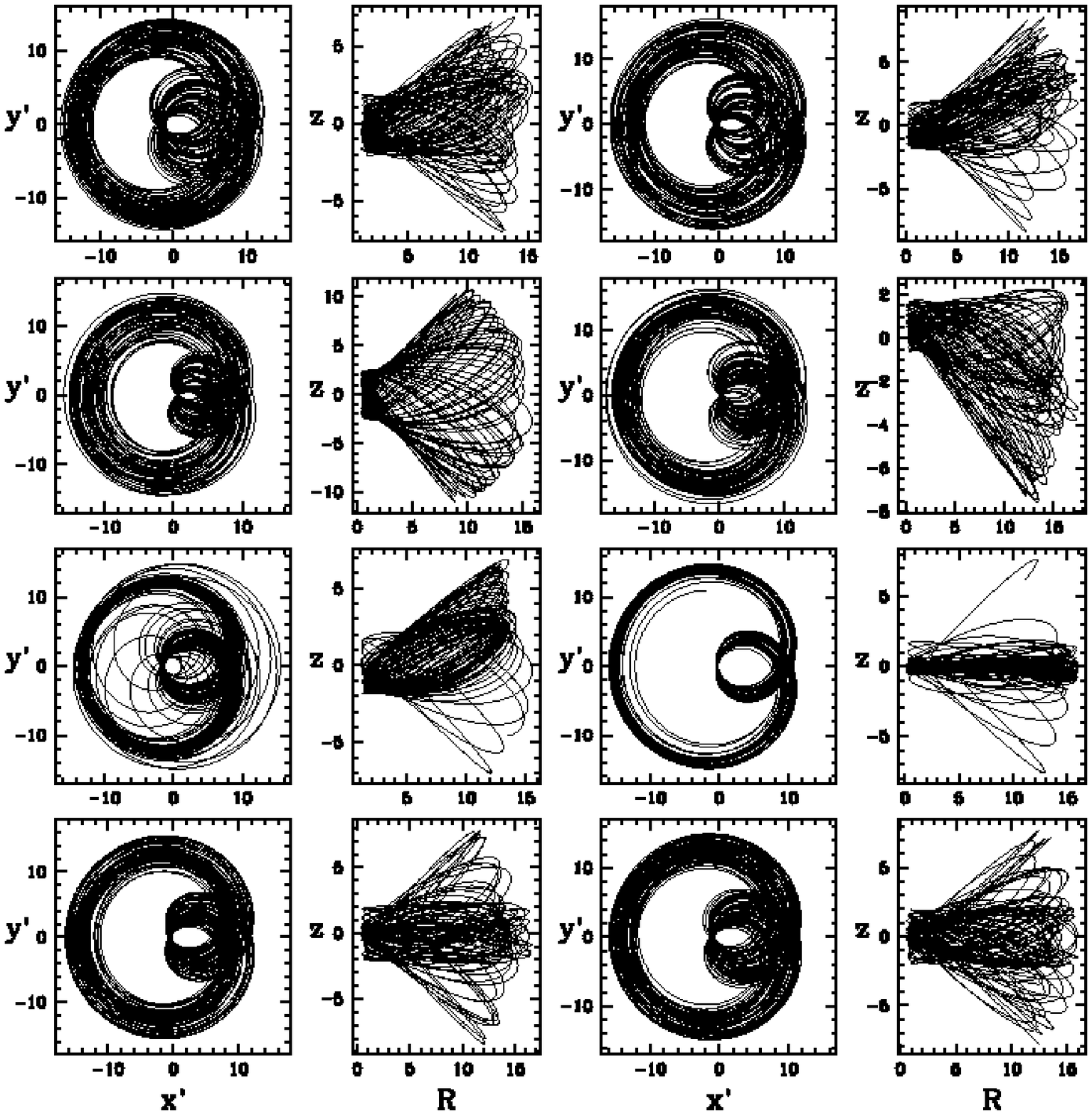}
  \caption{As in Figure \ref{figura29}, here for a sample of stars in
   region E in Figure \ref{figura21}.}
  \label{figura30}
\end{figure}

\begin{figure}
  \includegraphics[width=0.95\hsize]{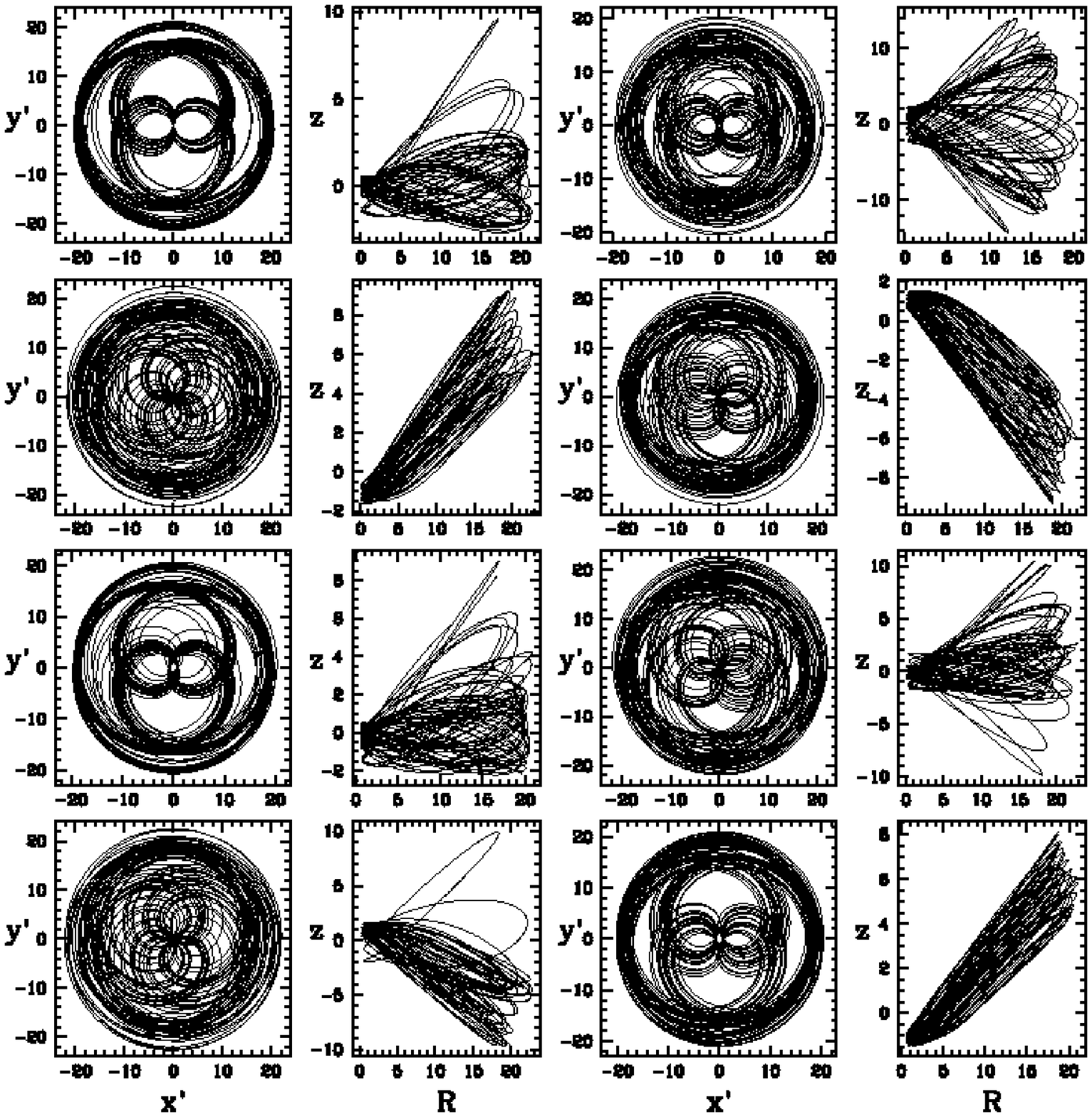}
  \caption{As in Figure \ref{figura29}, here for a sample of stars in 
   region F in Figure \ref{figura21}.}
  \label{figura31}
\end{figure}

For a sample of stars lying in regions D,E,F and reaching high
$z$-distances, Figures \ref{figura29},\ref{figura30},\ref{figura31},
respectively, show eight pairs of frames, four of them arranged in the
first and second columns and the other four in the third and fourth
columns. In each pair, the frame at the left gives the orbital
projection on the Galactic plane x$'$,y$'$ in the reference frame
where the bar is at rest, and the corresponding meridional orbit in
the plane $R,z$ is shown in the frame at the right; $R$ is the
distance from the Galactic rotational axis $z$.  As commented above,
stars in these regions D,E,F are mainly associated with families
V,IX,XI, respectively. This can be noticed comparing the orbital
projections on the Galactic plane in these figures with the form of
the periodic orbits in families V,IX,XI (see Figures
\ref{fig:famper5}, \ref{fig:famper9}, \ref{fig:famper11}).  Figure
\ref{figura29} shows in the third pair of frames at the left side an
orbit which seems to be associated with the family IV, and Figure
\ref{figura31} shows in the second and fourth pairs of frames at the
left side two examples of orbits which seem to be associated with the
family X, but their projections on the Galactic plane have departed
considerably from the unstable periodic orbit in this family X.  The
conclusion from Figures \ref{figura29},\ref{figura30},\ref{figura31}
is that some resonances $on$ $the$ $Galactic$ $plane$ can trap stars
reaching high $z$-distances from this plane. In their $N$-body models
of barred galaxies, \citet{CK07} found that particles in the halo with
a $z$-distance lower that 3 kpc could be trapped by the corotation and
inner Lindblad resonances. In our case, for the analysed resonances,
the trapping $z$-distance can be larger.

\begin{figure}
  \includegraphics[width=0.95\hsize]{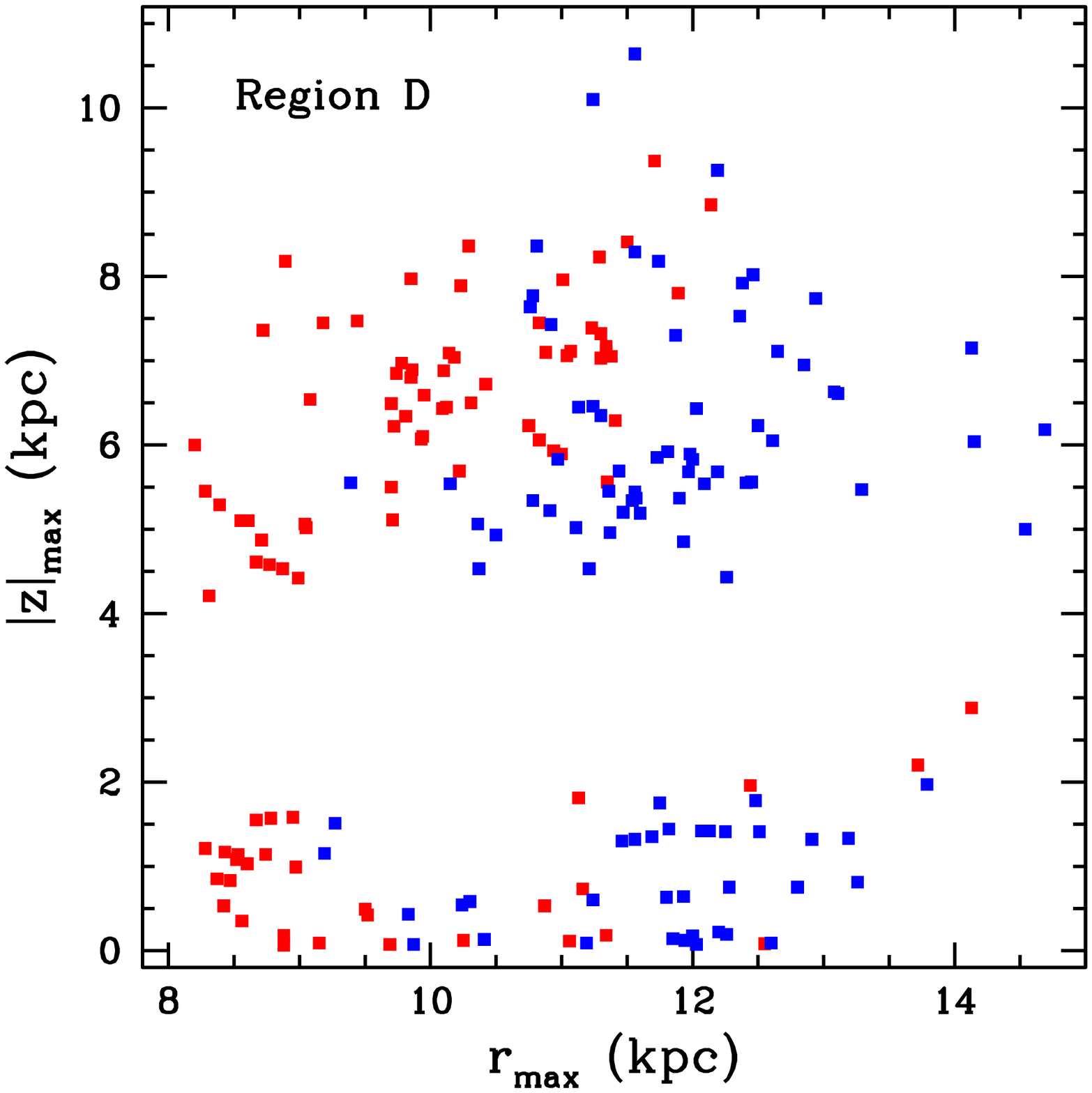}
  \caption{For member stars in the region D, this figure shows in the
  axisymmetric potential (red points) and nonaxisymmetric potential
  (blue points) the maximum orbital $z$-distance versus $r_{\rm max}$,
  with $r_{\rm max}$ the maximum distance from the Galactic centre.
  The separation in $|z|_{\rm max}$ is present in both potentials.}
  \label{figura32}
\end{figure}

\begin{figure}
  \includegraphics[width=0.95\hsize]{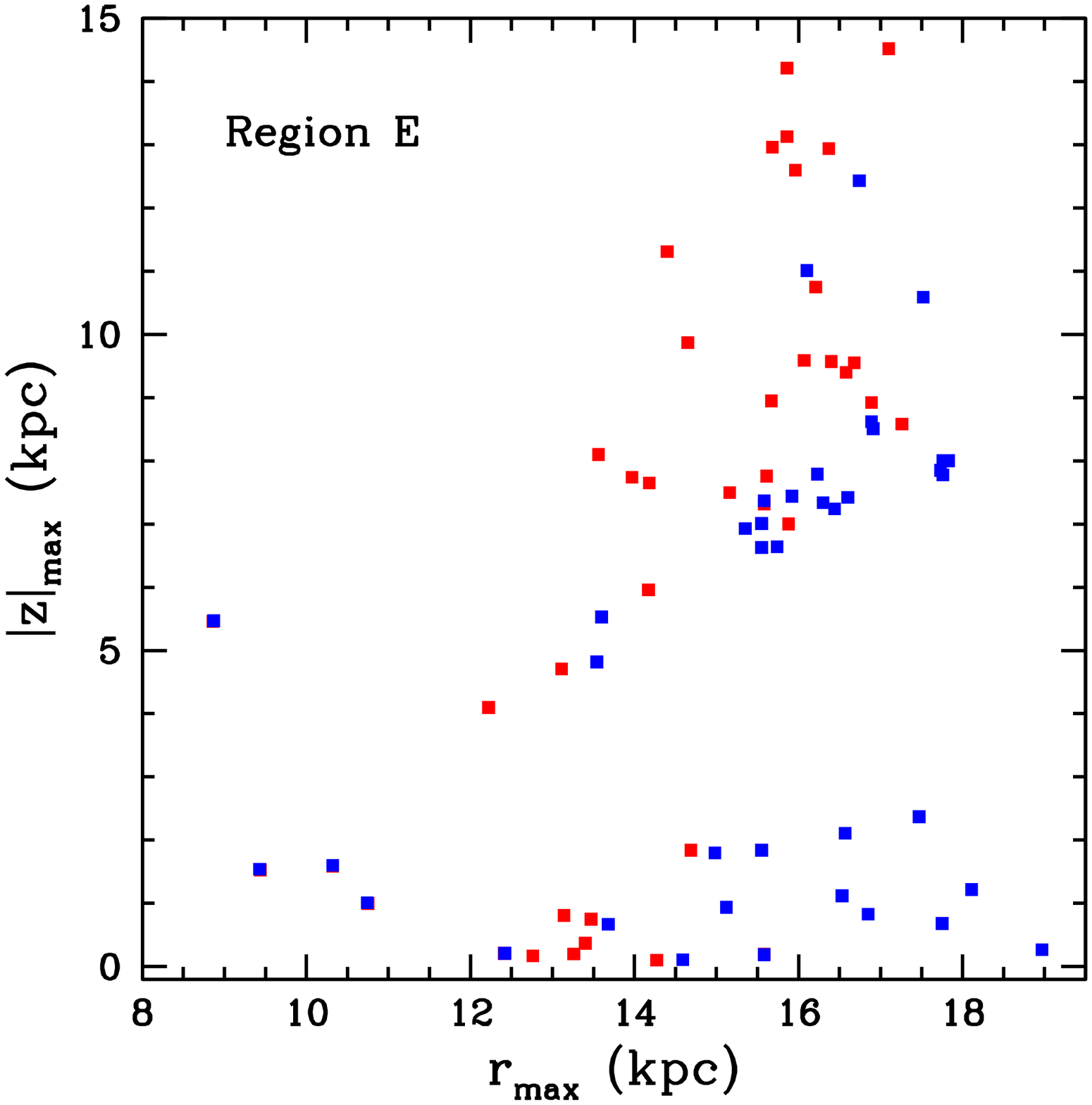}
  \caption{As in Figure \ref{figura32}, here for member stars in
  the region E.} 
  \label{figura33} 
\end{figure}

\begin{figure}
  \includegraphics[width=0.95\hsize]{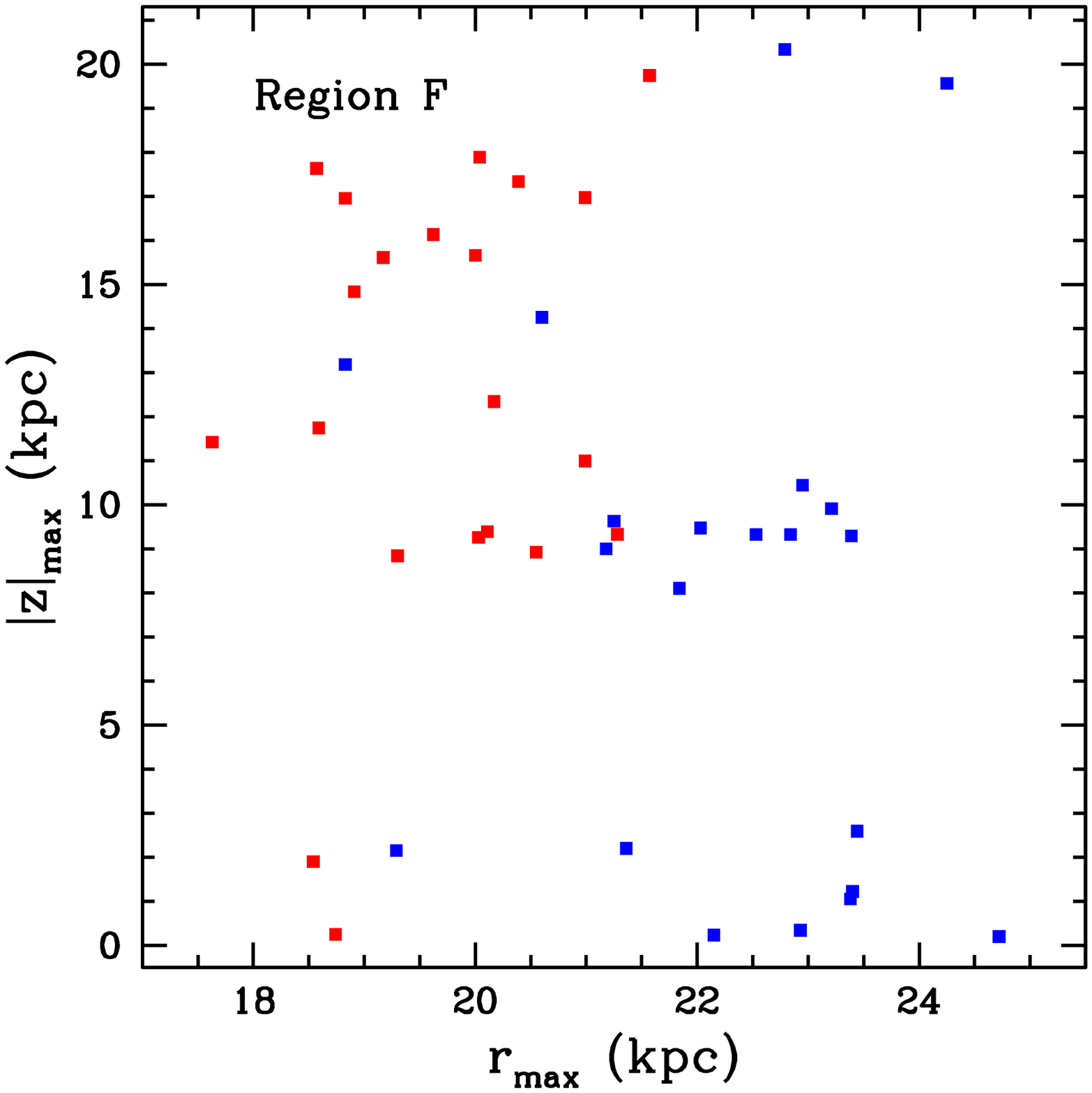}
  \caption{As in Figure \ref{figura32}, here for member stars in 
  the region F.} 
  \label{figura34}
\end{figure}

As it is shown in Figures \ref{figura25} to \ref{figura28}, for the
regions C,D,E,F there is some internal separation of points for the
$|z|_{\rm max}$-distance.  This separation is also present when we
compute the orbits in the axisymmetric potential obtained by
transforming the bar back into the original spherical bulge (see
Section \ref{sec:model}). Figures \ref{figura32} to \ref{figura34}
show the situation in the regions D,E,F. In these figures we plot
$|z|_{\rm max}$ versus $r_{\rm max}$, with $r_{\rm max}$ the maximum
orbital distance from the Galactic centre. The red points are obtained
with the axisymmetric potential and the blue points with the
nonaxisymmetric potential.

\begin{figure}
  \includegraphics[width=0.95\hsize]{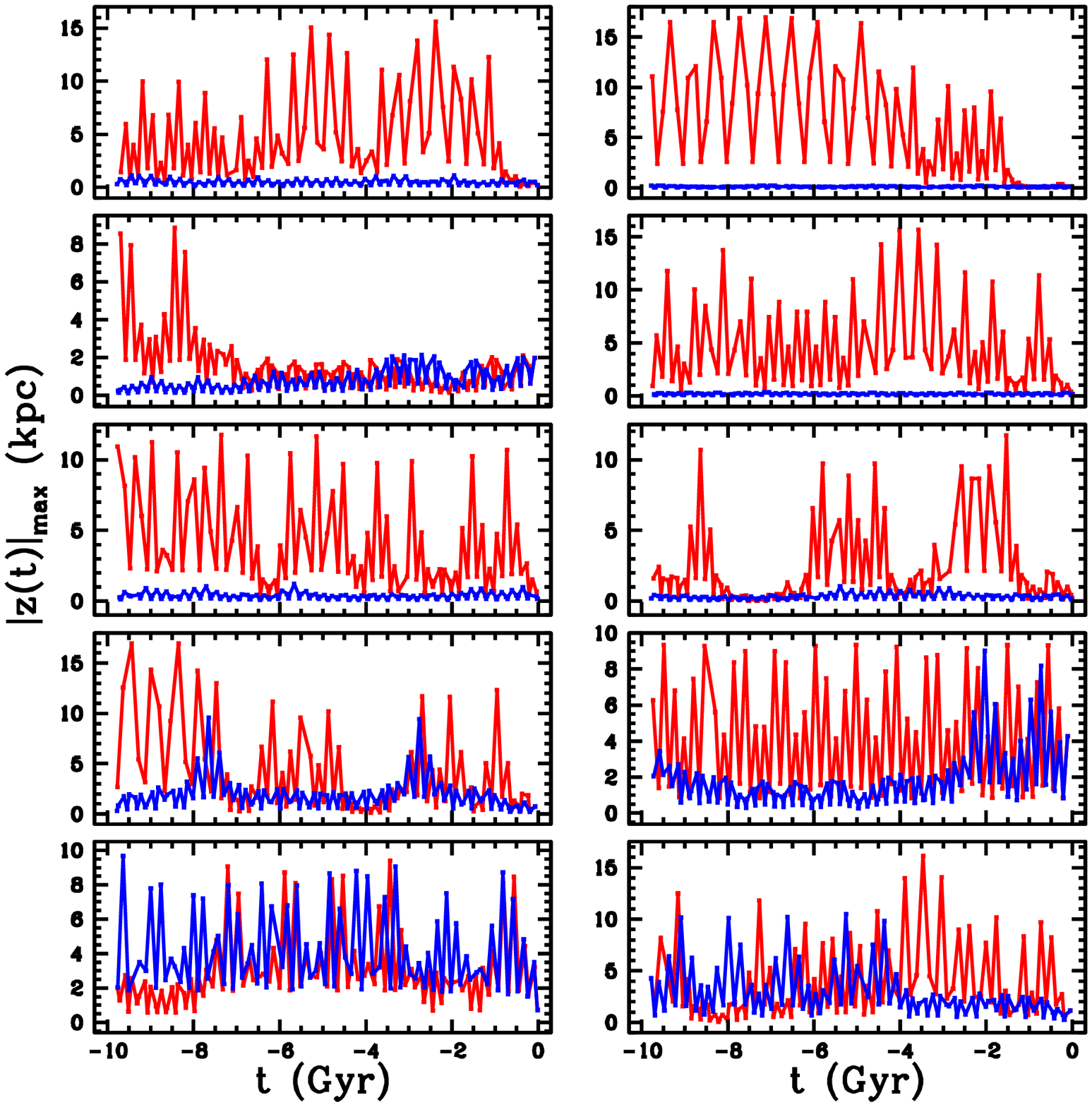}
  \caption{Maximum orbital $|z|$-amplitude as a function of time for
    ten member stars in the region F. The red and blue curves show the
    results in the axisymmetric and nonaxisymmetric potentials,
    respectively.}
  \label{figura35}
\end{figure}

The separation in $|z|_{\rm max}$ in the regions C,D,E,F seems to be
associated with an orbital transition between low- and high-$|z|$
orbital domains. In Figure \ref{figura35} we show how the maximum
$|z|$-amplitude varies in time for ten member stars in the region F;
the present time is at t = 0. The red curves show the results with the
axisymmetric potential and the blue curves the corresponding in the
nonaxisymmetric potential. The first three rows in the figure show six
orbits with points in the high-$|z|_{\rm max}$ domain in the
axisymmetric potential, but with points in the low-$|z|_{\rm max}$
domain in the nonaxisymmetric potential. Notice that for the first
star in the second row its orbit has suffered a transition to the
low-$|z|_{\rm max}$ domain, but for the whole computed time of 10 Gyr
it has been assigned to the high-$|z|_{\rm max}$ domain. The last two
rows in the figure also show orbital transitions in the axisymmetric
and nonaxisymmetric potentials.  In particular for stars which can be
trapped by the resonance associated to the region F, Figure
\ref{figura35} shows that the presence of the Galactic bar might favor
the motion close to the Galactic plane, compared with that obtained in
the axisymmetric potential; this result is also present in the
resonance associated to the region E, see Figure \ref{figura33}.  The
separations in $|z|_{\rm max}$ might be explained by the existence of
a major 1:1 resonant family of tube orbits which passes a few kpc
above the Sun and which can be appreciated in a number of the
`vertical' Poincar\'e diagrams given in \citet{SA97}, 
such as figs. 3a, 5a, and 6a.  Such a family limits the phase space
available for local halo-star orbits, and non-members of these tube
orbits must navigate above or below the resonant islands visible in
these `vertical' Poincar\'e diagrams, leading to the separations
visible in Figs. \ref{figura25} to \ref{figura28} and \ref{figura32}
to \ref{figura34}, and in the behaviour of Fig. \ref{figura25}.  Also,
stars which belong to such a 1:1 resonant family are not found in the
solar neighbourhood, but further above or below the Galactic plane
than the stars of the observational sample described in Section
\ref{catalogo}, maintaining these empty separations. This issue, and
its possible consequences for the external shape of the Galactic halo,
needs further study, analysing the resonant families external to the
region F to see if they produce a similar behaviour.

This section has shown that resonances on the Galactic plane can have
a strong influence on stellar orbits reaching high-$z$ distances from
this plane.  The stars of the Galactic halo may be strongly affected,
acting as test particles influenced by massive nonaxisymmetric
structures of the Galactic disc and bar.  This effect may have
important consequences in the formation of moving groups in the halo
associated with periodic orbits on the Galactic plane.

For stars with high-$z$ distances, there are other factors that may
influence their stellar kinematics, such as the triaxiality of the
Galactic halo \citep{PWG09,RVP12,RMP15}, that with the same resonant
process as the nonaxisymmetric bar structures, trigger the production
and support of stellar moving groups.  The relative importance of the
kinematics imprinted by resonances of the nonaxisymmetric bar
structure versus the one imprinted by the triaxiality of the halo is a
study that will be presented in a future paper.

\section{Spectral Analysis}\label{fourier}

To complement the analysis presented in previous sections, a spectral
analysis of the three-dimensional orbits of the 1642-star sample has
been made, and also for the two-dimensional orbits of the several
families of periodic orbits on the Galactic plane computed in Section
\ref{periodicas}. We have considered the method given by
\citet{PTVF92} for the spectral analysis of unevenly sampled data, and
their fast subroutine $fasper$ has been employed. In a given orbit,
computed in the noninertial reference frame where the bar is at rest,
the spectral analysis was made for each Cartesian coordinate x$'$(t),
y$'$(t), z$'$(t), or only x$'$(t), y$'$(t) in the case of the families
of periodic orbits. The dominant frequency (time$^{-1}$) was obtained
in each coordinate, i.e. the one with the largest amplitude or power
in the spectrum.

\begin{figure}
  \includegraphics[width=0.95\hsize]{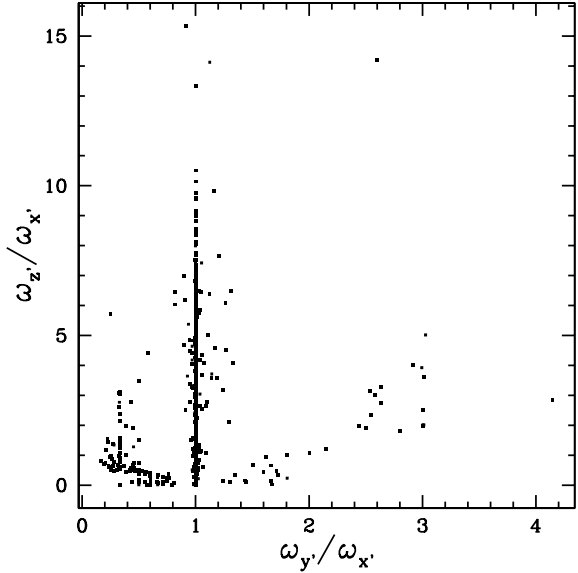}
  \caption{Ratios of the dominant orbital angular frequencies in the
  three Cartesian coordinates x$'$(t),y$'$(t),z$'$(t) for the 1642
  sample stars.}
  \label{figura36}
\end{figure}

For the 1642-star sample, Figure \ref{figura36} compares the ratios of
the dominant angular frequencies ${\omega}_{x'}$, ${\omega}_{y'}$,
${\omega}_{z'}$ (this figure is similar to figure 3.45 in
\citet{BT08}). The majority of the sample stars distributes around
${\omega}_{y'}/{\omega}_{x'}$ = 1, or with the resonant condition
      {\boldmath ${\omega}$}$\cdot$$\bf n$ = 0, with $\bf
      n$=(1,$-$1,0). Other populated parts appear at the lower region
      in the figure, especially at the lower left with $\bf
      n$=(1,$-$3,0) and in the diagonal line $\bf n$=(1,$-$1,$-$1).
      At the right side in this figure there is a concentration around
      $\bf n$=(3,$-$1,0).  The lower region in this diagram
      corresponds to halo stars, in which ${\omega}_{z'}$ decreases
      (the orbital period in the z-direction increases).

\begin{figure}
  \includegraphics[width=0.95\hsize]{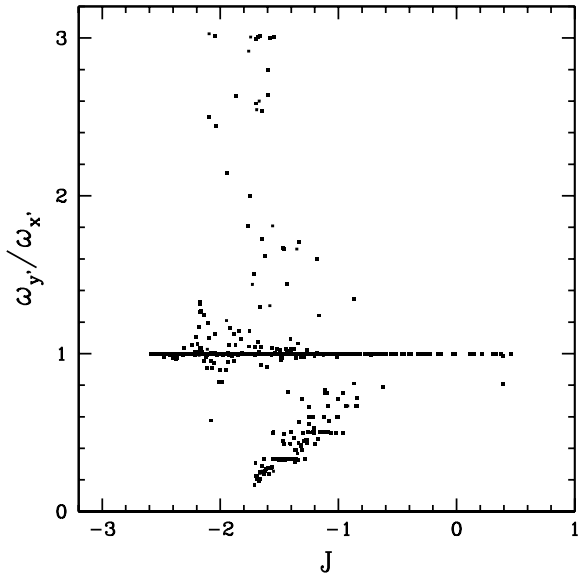}
  \caption{The ratio ${\omega}_{y'}/{\omega}_{x'}$ as a function of
  $J$ for the 1642-star sample.  Here we take $J \leq 1$ as in the
  zoom shown in Figure \ref{figura5}.} 
  \label{figura37}
\end{figure}

To make a diagram similar to those constructed in Section
\ref{sec:method}, in Figure \ref{figura37} we plot for the 1642-star
sample the ratio ${\omega}_{y'}/{\omega}_{x'}$ versus the corresponding
orbital Jacobi constant $J$ (taking $J \leq 1$ as in Figure
\ref{figura5}).  Again, several agglomerations of points are obtained,
with particular details in the interval $\approx$ ($-$2,$-$0.5) in $J$,
where regions D,E,F in Figure \ref{figura21} are located. 

\begin{figure}
  \includegraphics[width=0.95\hsize]{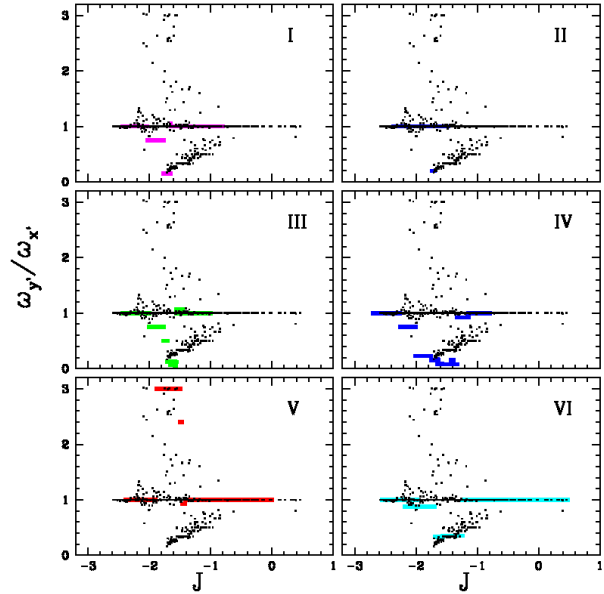}
  \caption{The ratio ${\omega}_{y'}/{\omega}_{x'}$ as a function of
  $J$ for families of periodic orbits I to VI. The colour of lines in
  each family is that employed in Figure \ref{figura6}. The black
  points correspond to the 1642-star sample.}
  \label{figura38}
\end{figure}

\begin{figure}
  \includegraphics[width=0.95\hsize]{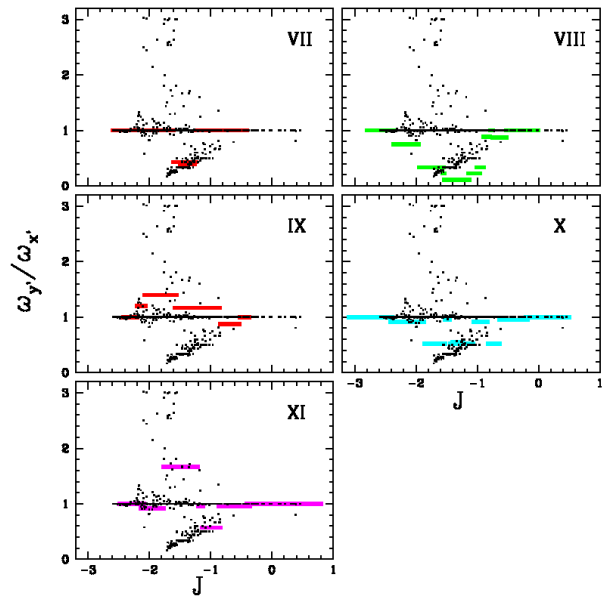}
  \caption{As in Figure \ref{figura38}, here for families VII to XI.}
  \label{figura39}
\end{figure}

The next step is to plot in a diagram like the one shown in Figure
\ref{figura37}, the several families of periodic orbits computed in 
Section \ref{periodicas}.  Figures \ref{figura38} and \ref{figura39} 
show this diagram separately for each family; the black points in each
frame correspond to the sample stars and the coloured lines to the
given family. The number of the family is given in each frame and the
employed colour is the same as in Figure \ref{figura6}. Notice that the 
ratio ${\omega}_{y'}/{\omega}_{x'}$ changes in a given family as $J$ is
varied (when there is a transition between two levels in this ratio,
its value may oscillate between the two levels and this appears as an
overlapping of levels at the transition). The agglomerations of sample
stars, which have three-dimensional orbits, are nearly covered by the
families of periodic orbits, which lie $on$ $the$ $Galactic$ $plane$.

\begin{figure}
  \includegraphics[width=0.95\hsize]{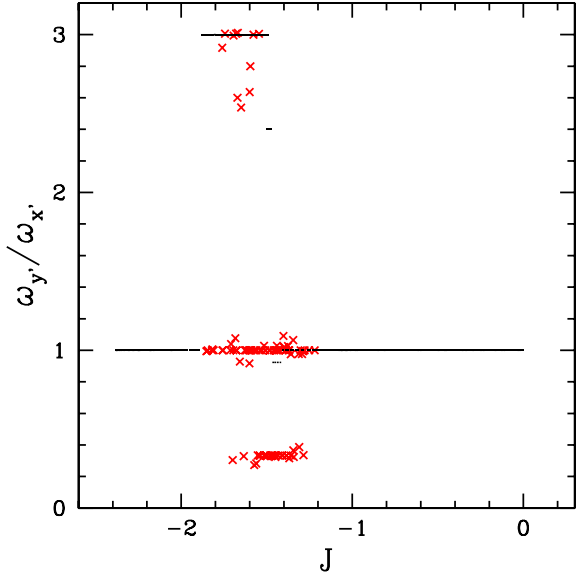}
  \caption{Comparison in a diagram ${\omega}_{y'}/{\omega}_{x'}$
  versus $J$ between the family V of periodic orbits (black lines) and
  the region D in Figure \ref{figura21} (red points).}
  \label{figura40}
\end{figure}

\begin{figure}
  \includegraphics[width=0.95\hsize]{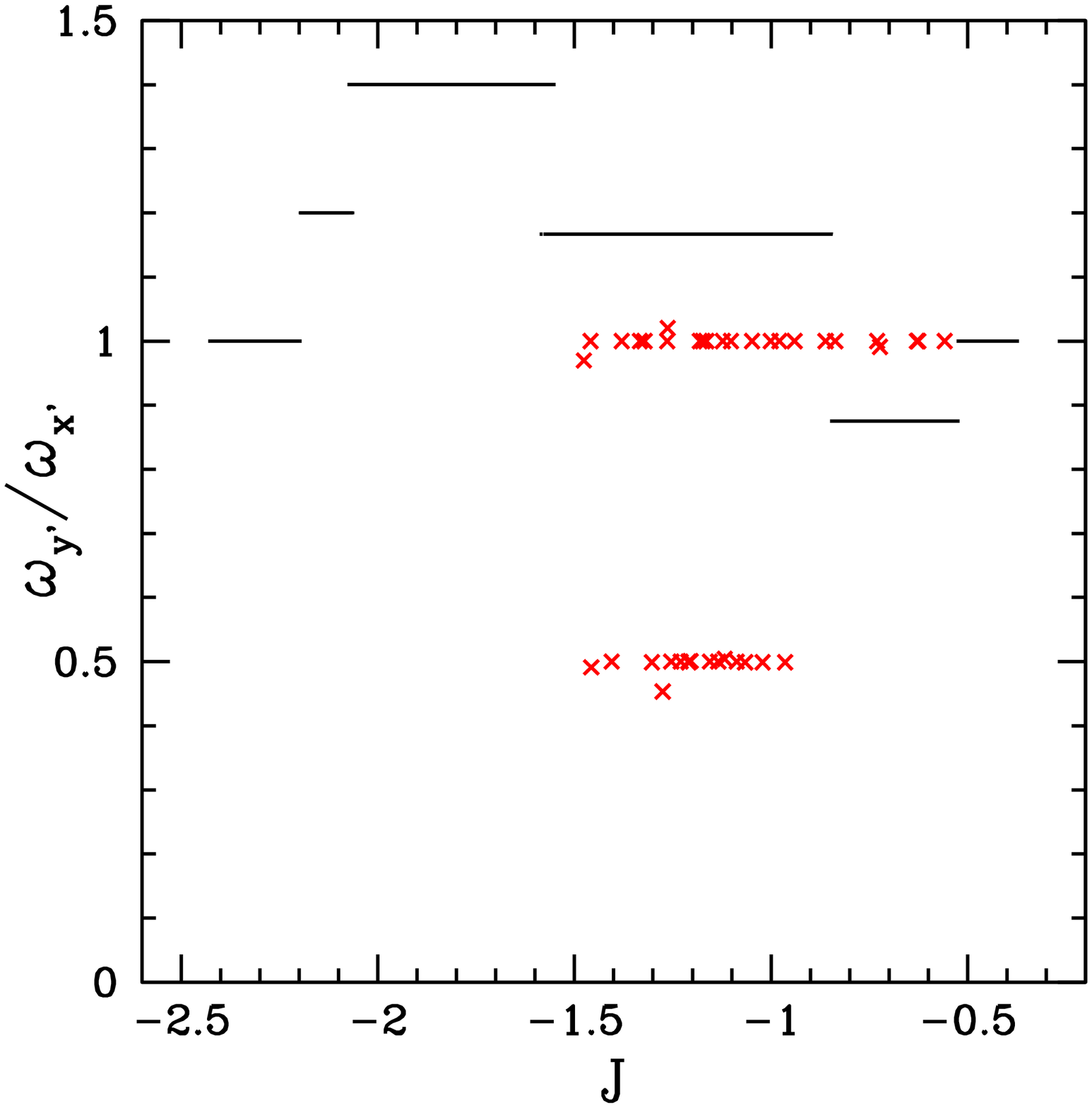}
  \caption{Same as in Figure \ref{figura40}, here for the family IX of
  periodic orbits and the region E in Figure \ref{figura21}.} 
  \label{figura41}
\end{figure}

\begin{figure}
  \includegraphics[width=0.95\hsize]{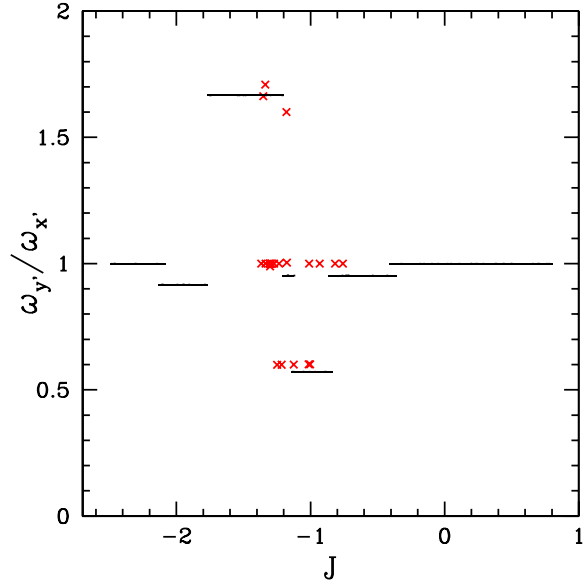}
  \caption{Same as in Figure \ref{figura40}, here for the family XI of 
  periodic orbits and the region F in Figure \ref{figura21}.}               
  \label{figura42}
\end{figure}

Now, in particular, the halo regions D,E,F shown in Figure
\ref{figura21} are considered.
We have seen that these regions are likely associated
with families V,IX,XI, respectively, which have stable parts in the
positions of these regions.  Thus, it is of interest to see what is the
relation region-family:  D-V, E-IX, and F-XI in a diagram
${\omega}_{y'}/{\omega}_{x'}$ versus $J$.  Figures \ref{figura40},
\ref{figura41}, and \ref{figura42} show the results. The black
lines correspond to the family and the red points to the region.
These figures show that there are approximate relations between
the corresponding region and family. A three-dimensional orbit which
may be trapped by a resonant family on the Galactic plane, will have a
tube orbital projection on this plane which resembles the corresponding
periodic orbit in the family. Only if this tube orbit is sufficiently
narrow, the dominant frequencies in the x$'$ and y$'$ directions will
be similar in both the projected and periodic orbits. The examples
given in Figures \ref{figura29}, \ref{figura30}, and \ref{figura31}
show that some resulting tube orbital projections of three-dimensional
orbits in regions D,E,F may have an appreciable thickness; thus, their
dominat frequencies may differ from those obtained in the resonant
family. This is shown explicitly in the next three figures 43, 44, and
45 commented below. The conclusion from this analysis of frequencies
is that a diagram ${\omega}_{y'}/{\omega}_{x'}$ versus $J$ may not give
a complete answer concerning the trapping of a three-dimensional orbit
by a resonant family on the Galactic plane; it is necessary to
complement the analysis with figures showing the orbital projection
on this plane, and include other frequencies in the spectrum, as
commented in the following. As concluded in the previous section, and
complemented with the analysis in this section, the halo regions
D,E,F have stars that are trapped by resonant families on the
Galactic plane.

\begin{figure}
  \includegraphics[width=0.95\hsize]{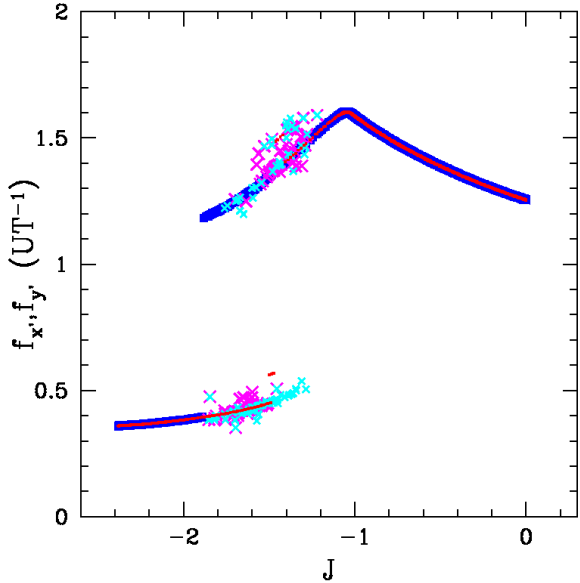}
  \caption{The frequencies f$_{x'}$, f$_{y'}$ in (UT)$^{-1}$, with 1
    UT = 0.1 kpc$\,$km$^{-1}$$\,$sec, for the family V of periodic
    orbits and the region D in Figure \ref{figura21}. The red and blue
    lines show respectively the values of f$_{x'}$,f$_{y'}$ for the
    family V and the points with magenta and cyan colours show
    respectively the values of f$_{x'}$,f$_{y'}$ for the region D.}
  \label{figura43}
\end{figure}

\begin{figure}
  \includegraphics[width=0.95\hsize]{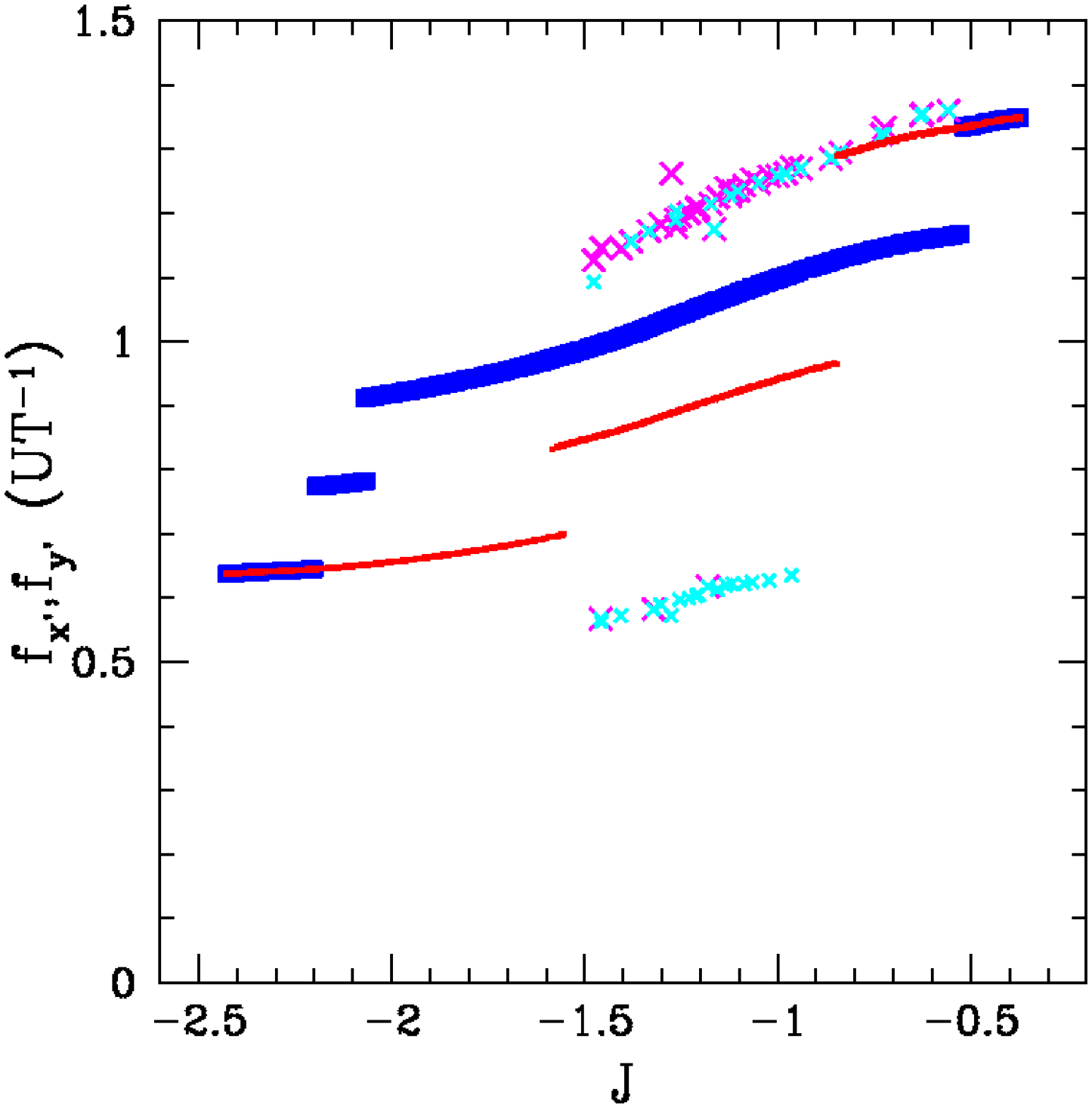}
  \caption{As in Figure \ref{figura43}, here for the family IX and
  the region E.}
  \label{figura44}
\end{figure}

\begin{figure}
  \includegraphics[width=0.95\hsize]{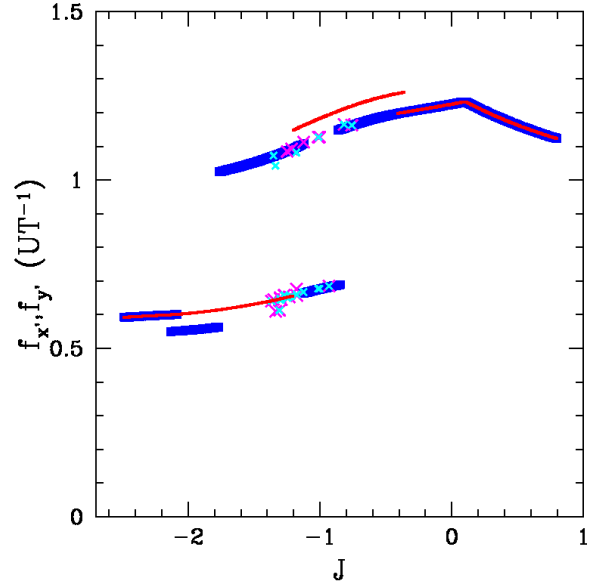}
  \caption{As in Figure \ref{figura43}, here for the family XI and
  the region F.}
  \label{figura45}
\end{figure}

An alternate form to present the results in Figures \ref{figura40},
\ref{figura41}, and \ref{figura42} is to plot directly the dominant
frequencies f$_{x'}$, f$_{y'}$ in the x$'$ and y$'$ coordinates versus
$J$. This is shown in Figures \ref{figura43}, \ref{figura44}, and
\ref{figura45}. The frequencies are given in (unit of time)$^{-1}$ =
(UT)$^{-1}$, with 1 UT = 0.1 kpc$\,$km$^{-1}$$\,$sec = 9.784 $\times$
10$^7$ yr. In these figures the red and blue lines show respectively
the values of f$_{x'}$ and f$_{y'}$ for the given family of periodic
orbits. The points (i.e. crosses) with magenta and cyan colours show
the corresponding values of f$_{x'}$,f$_{y'}$ for stars in the given
region. As commented above, and explicitly shown in these figures,
f$_{x'}$,f$_{y'}$ for a three-dimensional orbit in a given region may
not coincide with those obtained in the associated resonant family.
A more complete comparison would be obtained with these type
of figures by plotting not only the dominant frequencies in each case,
but also other peak frequencies in the corresponding spectrum,
especially if there are peaks with power similar to that obtained in
the dominant frequency and thus not considered in the previous
figures.

\section{Application of the Method}\label{ejemplo}

To illustrate the possible relation between the agglomerations of
stars trapped by resonances obtained in the empirical diagram of
characteristic energy versus $J$ presented in Section \ref{sec:method}
and groups of stars catalogued as moving groups, the position of the
Kapteyn group, a known moving group in the Galactic halo, is
considered for this diagram. The 1642 halo stars and disc stars in the
solar neighbourhood presented in Section \ref{catalogo} are again
employed to delineate this diagram; they serve as a reference to
compare with the corresponding positions in the diagram of the sample
stars in the Kapteyn group. In fact, some of these 1642 stars are
members of this moving group.

Eighteen member stars of the Kapteyn moving group have been selected
from \citet{E65a,E77,E90,E96a,E96b} and \citet{WFW10}.  The positions, 
proper motions, radial velocities, and distances of these stars have
been improved as was performed for the sample of 1642 stars.

\begin{figure}
  \includegraphics[width=0.95\hsize]{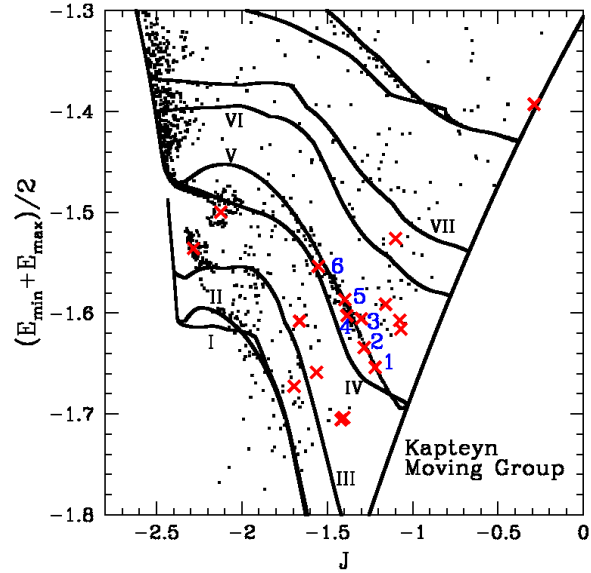}
  \caption{Diagram of characteristic energy versus $J$ in the
    nonaxisymmetric Galactic potential with the bar, showing with red
    points the positions of 18 members of the Kapteyn moving
    group. The small black dots correspond to the sample stars
    employed in Section \ref{sec:method}; the black curves are some
    families of periodic orbits given in Section \ref{periodicas}. The
    orbits of points (stars) marked as 1 to 6 are presented in Figures
    \ref{figura47}, and \ref{figura48}.}
  \label{figura46}
\end{figure}

Figure \ref{figura46} shows the diagram with the observational points
employed in Section \ref{sec:method} and some curves of families of
periodic orbits of Section \ref{periodicas}, along with red points
corresponding to stellar members of the Kapteyn moving
group. Comparing with Figure \ref{figura6}, these members arrange
mainly around families III, IV, and V. However, only the family V has
a stable part in its corresponding region covered by these member
stars; thus, the points marked as 1 to 6 in Figure \ref{figura46}
might have at the present time a relation with this resonant family V.

\begin{figure}
  \includegraphics[width=0.95\hsize]{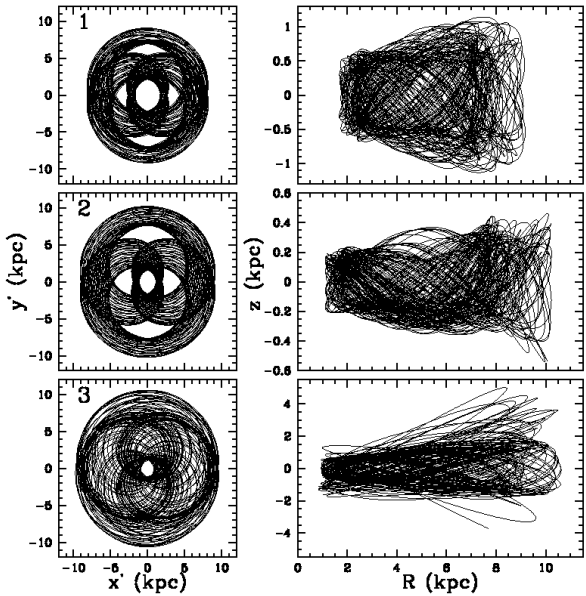}
  \caption{For the three points marked as 1,2,3 in Figure
   \ref{figura46}, the frames on the left show the corresponding 
   orbital projections on the Galactic plane in the x$'$,y$'$ axes,
   and the frames on the right show the meridional orbits.}
  \label{figura47}
\end{figure}

\begin{figure}
  \includegraphics[width=0.95\hsize]{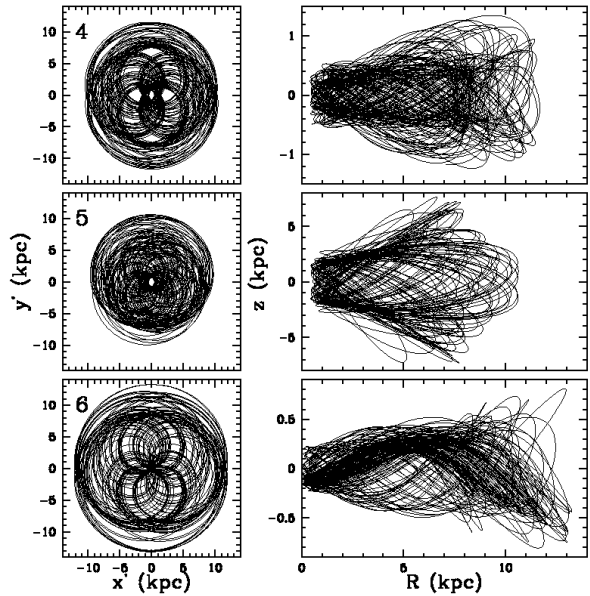}
  \caption{As in Figure \ref{figura47}, here for the three points
   marked as 4,5,6 in Figure \ref{figura46}.}
  \label{figura48}
\end{figure}

To see if this relation exists, for the six points (stars) marked as 1
to 6 in Figure \ref{figura46}, members of the Kapteyn moving group,
Figures \ref{figura47} and \ref{figura48} show at the left side their
corresponding orbital projections on the Galactic plane, plotted in
the x$'$,y$'$ frame where the bar is at rest. Their meridional orbits
are shown on the right side.  Points 1,2,4 have an orbital projection
on the Galactic plane which is a tube orbit, of the type that would be
obtained around stable periodic orbits in the family V (see panel 2 of
Figure \ref{fig:famper5}), in accordance with their positions on this
family, as shown in Figure \ref{figura46}. On the other hand, points 3
and 6 are very close to the curve of the family V, but their orbital
projections on the Galactic plane resemble an orbit departing from the
unstable periodic orbits in the family IV (see frame 3 in Figure
\ref{fig:famper4}). The orbit corresponding to point 5 does not seem
to be associated with any resonance. Thus, the majority of the member
stars of the Kapteyn moving group does not appear to have a relation
with a resonant family on the Galactic plane.

On the other hand, a direct connection between the Kapteyn moving
group and the well studied globular cluster $\omega$ Centauri has been
argued by \citet{WFW10} using detailed kinematical and
chemical analyses.  They propose that many, or all, members of this
moving group have been left behind as remnants as this globular
cluster (ex-dwarf galaxy) was dragged into the Galaxy by dynamical
friction. \citet{MN05} have carried out numerical simulations
of such a process, and find a local group of stars which seem to
relate to $\omega$ Cen both in their angular momenta and in their
chemical abundance characteristics. \citet{SSC12} use the
Bottlinger, Toomre and [Fe/H],V(rot) diagrams to identify in the solar
neighbourhood halo moving-group candidates relating to both the
Kapteyn and the $\omega$ Cen entities. \citet{SM12} 
conclude that $\omega$ Cen may have contributed in a significant way
to their `low-alpha' local halo component.  If the Kapteyn group is
of `cosmological' origin, i.e. all stars having formed together in
$\omega$ Cen, then the appearance of Figure 46 may be due to the
migration of the Kapteyn stars to the nearby resonant families III,
IV, and V during their trajectories while being lost to the Galaxy
from $\omega$ Cen.  That is, Figure 46 seems to suggest a possible
secular origin for the Kapteyn moving group, i.e. a product of the
resonant trapping by the Galactic bar, but this is misleading, as
shown here.

In a second paper, other moving groups in the Galactic halo will be
analysed using the diagram of characteristic energy versus $J$. 
The method will also be applied to some known moving groups in the
the Galactic disc.

\section{Conclusions}\label{concl}

In this study we have analysed the trapping of stars by resonances 
on the Galactic plane created by the Galactic bar.  Our aim is
to relate this mechanism with moving groups in our Galaxy, especially
with moving groups in the Galactic halo.  A new method is presented to
delineate the trapping regions. To show how the method works, a sample
of halo stars and disc stars in the solar neighbourhood has been
employed. The orbits of these sample stars are computed in a
nonaxisymmetric Galactic potential, and empirical diagrams are
constructed using some particular orbital properties. These diagrams 
plot a characteristic energy versus a characteristic angular momentum,
or a characteristic energy versus the orbital Jacobi constant, if the
Galactic potential employs only the bar as its nonaxisymmetric
component.

Some agglomerations of points in the disc and halo regions
are obtained in these diagrams, which coincide with some families of
periodic orbits on the Galactic plane, induced by the Galactic bar and
plotted in the same diagrams.  The influence of these resonant families 
can extend some kpc from the Galactic plane, explaining the observed
coincidence in the halo region.  Our future aim with this analysis is
to investigate if these agglomerations obtained in the disc and halo 
can be the sites where some known moving groups lie.

The proposed method has been illustrated here with the Kapteyn group,
a known moving group in the Galactic halo, resulting in that a
majority of member stars in this group does not have a relation with
any of the resonant families or periodic orbits obtained in the
employed Galactic potential, instead, we propose that these group
might rather have an origin on the dissolution of a dwarf galaxy or a
globular cluster. In a second paper the method will be applied to
other groups in the Galactic halo, including also some known moving
groups in the disc, to provide one piece of information on their
possible, or not, resonant origin.

%%%%%%%%%%%%%%%%%%%%%%%%%%%%%%%%%%%%%%%%%%%%%%%%%%%%%%%%%%%%%%%%%%%%%%%
\section*{Acknowledgements}

We thank the anonymous referee for a very careful review and several
excellent suggestions that greatly improved this work. Based upon
observations acquired at the Observatorio Astron\'omico Nacional in
the Sierra San Pedro M\'artir (OAN-SPM), Baja California, M\'exico. We
thank Agust\'in M\'arquez, Benjam\'in Hern\'andez, Sergio Silva, and
Maru Contreras for their contributions, and also the technical staff
at SPM, especially Gabriel Garc\'ia (deceased), Gustavo Melgoza, and
Felipe Montalvo. The authors acknowledge financial support from
UNAM-DGAPA PAPIIT IN114114 and IN103014 grants, and the CONACyT
(M\'exico) projects 27884-E, CB-2005-49434, and CB-2005-49002. We also
acknowledge the use of the SIMBAD data base at the CDS, Strasbourg,
France, and the ADS of the SAO/NASA.

%%%%%%%%%%%%%%%%%%%%%%%%%%%%%%%%%%%%%%%%%%%%%%%%%%%%%%%%%%%%%%%%%%%%%%%

\label{lastpage} 
\end{document}